\documentclass{jaa}
\usepackage{txfonts}
\usepackage{graphicx}
\newcommand{\gtsim}{\stackrel >{_\sim}}
\newcommand{\ltsim}{\stackrel <{_\sim}}

\begin{document}

\title[GMRT Galactic HI Absorption survey I]
{A High Galactic Latitude HI 21cm-line Absorption Survey using the GMRT:
                         I. Observations and Spectra }
\author[Mohan, Dwarakanath \& Srinivasan]
{Rekhesh Mohan\thanks{Currently at the Indian Institute of Astrophysics, 
                      Bangalore 560 034, India.}
       \thanks{e-mail:reks@iiap.res.in},
       K. S. Dwarakanath\thanks{e-mail:dwaraka@rri.res.in},
       \& G. Srinivasan\thanks{e-mail:srini@rri.res.in} \\
        Raman Research Institute, Bangalore 560 080, India }
\maketitle
\label{firstpage}

\begin{abstract}
We have used the Giant Meterwave Radio Telescope (GMRT) to measure the
Galactic HI 21-cm line absorption towards 102 extragalactic radio continuum
sources, located at high ($|b|~>~15^{\circ}$) Galactic latitudes. The
Declination coverage of the present survey is $\delta~>~\sim~-45^{\circ}$. With
a mean rms optical depth of $\sim0.003$, this is the most sensitive Galactic
HI 21-cm line absorption survey to date. To supplement the absorption data,
we have extracted the HI 21-cm line emission profiles towards these 102 lines
of sight from the Leiden Dwingeloo Survey of Galactic neutral hydrogen. We
have carried out a Gaussian fitting analysis to identify the discrete
absorption and emission components in these profiles. In
this paper, we present the spectra and the components. A subsequent paper will 
discuss the interpretation of these results.
\end{abstract}


\section{Introduction}
\label{sec:intro}
The distribution atomic hydrogen in the Galaxy and its physical properties
have been extensively studied.
Soon after the discovery of the HI 21-cm line, a number of single dish
HI surveys were conducted (See Burton, \nocite{wbb88}1988 for a useful
compilation of the early HI surveys).
The single dish HI absorption spectra are limited by the errors due to
the variation of HI emission intensity over the angular scales smaller than
the telescope beam.
Interferometric surveys are a better alternative for HI absorption studies.
An interferometer rejects the low spatial frequencies where HI emission is
dominant, resulting in reliable absorption spectra.
There have been a number of interferometric HI absorption studies,
often supplemented by single dish observations to measure the HI emission.

The early HI 21-cm line absorption and emission studies led to the emergence of
a global picture of the interstellar medium (Clark, Radhakrishnan \& Wilson,
\nocite{crw62}1962; Clark, \nocite{barry65}1965). These and later
studies paved way for the models of the interstellar medium of the Galaxy.
Field, Goldsmith \& Habbing  \nocite{fgh69}(1969) modeled the ISM as
cool dense concentrations of gas, often referred to as
``interstellar clouds'' (the Cold Neutral Medium or CNM) in pressure
equilibrium with a warmer intercloud medium (the Warm Neutral Medium or WNM).
While this initial model of the ISM has been refined considerably by
later studies (Wolfire et al,
\nocite{wolfire95}1995), the basic picture of the ISM with cold diffuse clouds
and a warmer intercloud medium has survived.
The spin temperature of the
cold clouds which constitute the CNM was estimated to be $\sim$80 K and
that of the WNM to be $\sim$8000 K. The cold clouds manifest
as narrow Gaussian features with typical velocity dispersions of a few km
s$^{-1}$ in both HI emission and absorption profiles. The intercloud medium,
on the other hand, is identified with broad Gaussian features with typical
velocity dispersions $\gtsim$10 km s$^{-1}$. The intercloud medium is usually
detected in emission,
since the HI absorption in the warm gas is very weak.
We summarize the results from some of the important surveys below.

Radhakrishnan et al (\nocite{rad72a}1972a, \nocite{rad72b}1972b) used the
Parkes Interferometer to study the HI absorption towards 35 extragalactic
radio sources, in the Galactic latitude range 6$^{\circ}$ $\le$ $|b|$
$\le$ 74$^{\circ}$ and 53 Galactic radio sources in the lower Galactic
latitudes, mostly located at $|b|$ $<$ 2$^{\circ}$. In addition, the Parkes
64 m telescope was used to obtain the HI emission towards all these
directions. The velocity resolution of these observations was 2.1 km s$^{-1}$.
The rms noise in the HI optical depth profiles varied from $\sim$0.01 to
$>$0.1, depending on the flux density of the background source. They
concluded that the HI absorption features are arising in discrete
concentrations of gas with a spin temperature in the range 60 -- 80 K.
They also derived the number density of these features to be $\sim$2.5 per Kpc
(Radhakrishnan \& Goss, \nocite{rg72}1972). The number of such concentrations
of gas for a given optical depth $\tau$ was $\propto$
$e^{-\tau}$. Lack of absorption in the intercloud medium, which was identified
as the wide ``shoulders'' in HI emission profiles, enabled them to put a limit
of $>$750 K for the spin temperature of this gas.

Dickey \& Benson \nocite{dickey82}(1982) used the NRAO 300 ft and 140 ft
telescopes as an interferometer to detect absorption
in the 21-cm line towards 64 radio continuum sources. The HI emission profiles
were obtained using the 300 ft telescope alone. These sources were both
Galactic and extragalactic and spread over a range of latitudes
0$^{\circ}$ $\ltsim|b|$ $\ltsim70^{\circ}$. They have produced HI absorption
profiles with an rms optical depth in the range $\sim$0.007 -- $\sim$0.16,
with a velocity resolution ranging from 1.3 $-$ 5.3 km s$^{-1}$.
Among the main results of this study was the realization that for lower
Galactic latitudes ($|b|$ $\ltsim$15$^{\circ}$), HI emission surveys using
single dish telescopes would miss a significant amount of gas ($\sim$40\%)
due to HI self absorption, wherein the cold HI gas in the foreground absorbs
the HI line emission from the background gas. They also found more HI gas with
lower spin temperature (100 $\ltsim$T$_{S}$ $\ltsim$150 K) at lower latitudes
($|b|$ $<$ 2$^{\circ}$) as compared to the previous HI line surveys. They 
concluded that such a behaviour is the result of velocity blending.

Mebold et al \nocite{egs81}(1981) carried out an HI absorption survey towards
69 sources in the range 0$^{\circ}$ $\ltsim|b|$ $\ltsim$80$^{\circ}$
using the NRAO 3-element interferometer. They obtained the corresponding
HI emission using the Effelsberg 100 m telescope, the 91 m Green Bank telescope
or the 64 m Parkes telescope. The velocity resolution of these profiles were
in the range 0.42 -- 3.3 km s$^{-1}$. These profiles, on an average, had an
rms in HI optical depth $\gtsim$0.05. They found most of the HI 
absorption features to have a spin temperature in the range 20 to 140 K. For 
the HI absorption data at $|b|$ $>$ 15$^{\circ}$, they found indications for a 
bimodal distribution in the radial velocity distribution of absorbing features.
However, since the number of components at higher radial velocities were small,
they were unable to study its significance.

Till recently, the Very Large Array (VLA) used to be the only instrument with a 
collecting area comparable with large single dish telescopes. The survey by 
Dickey et al.
\nocite{dickey83}(1983) using the VLA is limited to lower Galactic latitudes
($|b|$ $<$ 10$^{\circ}$). Moreover, the optical depth detection limit for
this survey is $\sim$0.1 (3$\sigma$). From the various HI absorption surveys
carried out so far (Radhakrishnan et al., \nocite{rad72a}1972a,
\nocite{rad72b}1972b; Mebold et al., \nocite{egs81}1981; Dickey \& Benson,
1982 \nocite{dickey82}; Dickey et al., \nocite{dickey83}1983), more than
600 absorption spectra
are available, but the optical depth detection limits of more than 75\% of
these are above 0.1. 
From the available results, the cloud population observed in HI absorption
seem to have a Gaussian random velocity distribution. The dispersion in the
random velocities of HI absorption features is $\sim$7 km s$^{-1}$
(Dickey \& Lockman, \nocite{araa90} 1990, and references therein).

The low optical depth regime of Galactic HI is largely unexplored except for
the HI absorption studies using the Arecibo reflector by Dickey et al
\nocite{dst78}(1978) and more recently by Heiles \& Troland \nocite{ht03a}(2003a, 
\nocite{ht03b}b). Dickey et al \nocite{dst78} measured HI absorption
and emission towards 27 extragalactic
radio continuum sources located at high and intermediate Galactic latitudes
($|b|$ $>$ 5$^{\circ}$). The rms optical depth in their spectra were typically
$\sim$0.005. In many of the profiles the systematics in the band dominate 
the noise in the spectrum. As we noted earlier,
these observations are not impervious to HI  emission fluctuations introducing
errors in the absorption profile (Dickey \& Lockman, \nocite{araa90}1990).
Radhakrishnan \& Goss \nocite{rg72}(1972) found the number of HI absorption
features for a given optical depth, $\tau$ to be proportional to $e^{-\tau}$.
However, the data by Dickey et al \nocite{dst78}(1978) as well as the later
survey using the Green Bank Interferometer (Mebold et al, \nocite{egs82}1982)
indicated that for $|b|$ $>$ 15$^{\circ}$ this dependence is steeper than
$e^{-\tau}$. This trend was explained as due to increase of low optical depth
features with increasing angular and velocity resolution
(Mebold et al, \nocite{egs82}1982).
Dickey et al \nocite{dst78}(1978) also noted that the velocity distribution of
HI absorption features is dependent on the optical depth. For the optically thin
clouds ($\tau$ $<$ 0.1), the velocity dispersion was $\sim$11 km s$^{-1}$,
whereas for the optically thick clouds ($\tau$ $>$ 0.1) this value is $\sim$6
km s$^{-1}$. Heiles \& Troland (\nocite{ht03a}2003a, \nocite{ht03b}b)
analysed HI absorption and emission profiles toward 79 lines
of sight. A good fraction of these directions (66 out of 79) were at Galactic
latitudes $|b|$ $>$ 10$^{\circ}$. They found evidence for an excess of low
column density ($N_{\scriptscriptstyle{\rm{HI}}}$ $<$ 5 $\times$ 10$^{19}$
cm$^{-2}$) CNM components.

\section{Motivations for the present survey}
\label{sec:intentions}
The prime motivation for the present survey was to obtain sensitive HI
absorption measurements at high Galactic latitudes and to study the random
velocity distribution of HI clouds. Although there exist
extensive data on Galactic HI absorption, there is a lack of sensitive HI
absorption studies. There are indications that the low optical depth features
($\tau$ $<$ 0.1) form a distinct class (Dickey et al, \nocite{dst78}1978;
Mebold et al \nocite{egs82}1982; Heiles \& Troland \nocite{ht03b}2003b), with
larger velocity dispersion. The dependence of HI column density and optical
depths of these clouds on their random velocities are not well studied. Our aim
was to investigate the nature of low optical depth HI features in the Galaxy
and to estimate their velocity distribution. One of the difficulties encountered
in studying discrete components in the HI 21-cm line profiles in the Galactic
plane is the plethora of absorption lines. Larger path length through the
disk of the Galaxy results in larger number of absorption components.
In such cases, the available
techniques often result in more than one possible solution for the parameters
of individual features. Moreover, the observed radial
velocities are usually the sum of components arising from random motion and the
differential rotation of the Galaxy. The distances to the absorbing
clouds are seldom known. Hence the systematic component in the observed radial
velocity arising from Galaxy's rotation is unknown. Therefore, the lower
Galactic latitudes are not suitable for a survey to search for low optical
depth components and to study the random velocity distribution of interstellar
clouds. We have chosen a lower cutoff of
15$^{\circ}$ for the Galactic latitude in our observations.

Most of the HI gas at higher latitudes is observed only in HI emission
(Dickey \& Lockman, \nocite{araa90} 1990, and references
therein). The HI emission profiles often show components at velocities that
cannot arise from Galactic rotation. The existing surveys of HI absorption
indicate that there is very little absorption in directions above a latitude of
$\sim$45$^{\circ}$, down to optical depths $\sim$0.01 (Dickey et al,
\nocite{dst78}1978). But, there are indications that HI gas layer of the
Galaxy extends to several kpc (Albert, \nocite{al83}1983; Lockman \& Gehman,
\nocite{lg91}(1991); Kalberla et al, \nocite{kalb98}1998).  A more sensitive
HI absorption search is required to understand the nature of this gas and to
characterize it.

We present here the HI absorption measurements with the Giant Metrewave
Radio Telescope (GMRT) towards 102
extragalactic radio continuum sources located at intermediate and high
latitudes. The present survey, with an rms detection limit of 0.003 in HI
optical depth is at least a factor of 5 more sensitive than
the existing interferometric surveys and is comparable with the 
sensitivities achieved in the single dish HI surveys using the Arecibo 
telescope (Dickey et al, \nocite{dst78}1978; Heiles \& Troland, 
\nocite{ht03a}2003a).
An overview of the GMRT is given in the next section. The strategy for
selecting the sources is outlined in section \ref{sec:sources} and the details
pertaining to the observations are given in section \ref{sec:obs}. A brief
description of the data analysis is given in section \ref{sec:data} and
section \ref{sec:spectra} describes a sample HI absorption
profile. The list of observed sources is presented in Appendix A, the HI line
profiles towards these sources are given in Appendix B and Appendix C lists
the discrete HI features identified from each of these profiles.

\section{The Giant Meterwave Radio Telescope}
\label{sec:gmrt}
The Giant Meterwave Radio Telescope (GMRT) consists of 30 fully steerable
dishes, of 45 meter diameter with a maximum baseline of 25 km (Swarup et al.,
\nocite{gmrt}1991). The aperture efficiency of the dishes is $\sim$ 40$\%$ in
the 21cm band, which implies an effective collecting area of $\sim$ 19000
m$^{2}$. The full width at half maximum of the primary beam primary beam is 
$\sim$25$^{\prime}$ and that of the
synthesized beam is $\sim$2$^{''}$ (uniform weighting) at 1.4 GHz.
The 21 cm receiver is a wide band system covering the frequency range 900 --
1450 MHz. It is a prime focus uncooled receiver with a characteristic
system temperature of $\sim$70K. The 21cm system has four sub bands, centered
at 1060, 1170, 1280 and 1390 MHz respectively, each with a 3 dB bandwidth of 
120 MHz. At the time when these observations were carried out, this telescope 
was equipped with an FX correlator providing 128 channels per polarization per
baseline. A baseband bandwidth ranging from 16 MHz down to 64 kHz variable in
steps of 2 can be chosen.

\section{Source selection}
\label{sec:sources}
In selecting the sources, we have used a lower cutoff in flux density of 1 Jy
at 20cm wavelength. This was to ensure that we reach an  rms in HI optical
depth $\sim$0.003 within an observing time $\ltsim$1 h.
In order to obtain a sample of these bright sources uniformly distributed all
over the sky we used the VLA calibrator
manual as the basic finding list. A list of 102 point sources,
unresolved in the VLA B array configuration, with their Galactic
latitude $|b|$ $\geq$ 15$^{\circ}$ were selected. Figure \ref{fig:sourcedist}
shows the distribution of the program sources in the sky. The list of sources
is given in Appendix A.

To supplement the absorption spectra, HI emission profiles were
extracted from the Leiden-Dwingeloo all sky survey of Galactic neutral hydrogen
(LDS, Hartmann \& Burton, \nocite{lds}1995). This survey used the 25 meter
Dwingeloo telescope to map the sky in HI emission. The full width at half maximum
of the primary beam of the Dwingeloo telescope is $\sim$36$^{'}$. The
geographic latitude of the Dwingeloo telescope is $\sim$+53$^{\circ}$ and that
of the GMRT is $\sim$+19$^{\circ}$. Therefore some of the lines of sight in
the southern hemisphere observed with the GMRT are not accessible to the
Dwingeloo telescope.

\begin{figure}
{\centering\includegraphics[width=7.5cm,height=13.0cm,angle=-90]{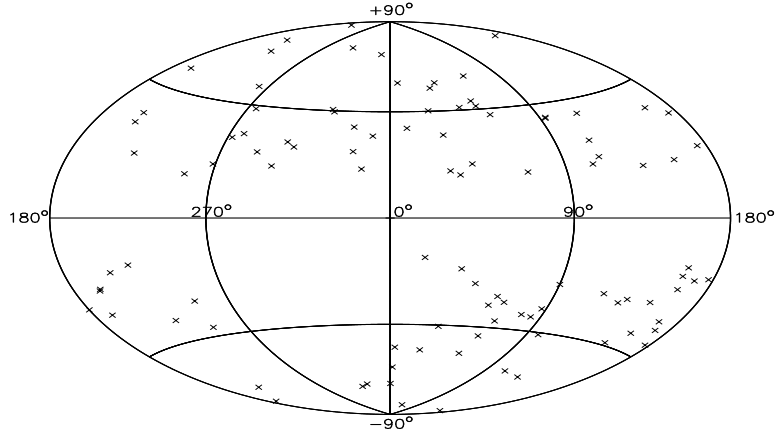}}
\caption{The distribution of program sources in Galactic co-ordinates. The
fourth Galactic quadrant (270$^{\circ}$ $<$ l $<$ 360$^{\circ}$) is not fully
accessible for the GMRT, since this region is mainly in the southern
equatorial hemisphere}
\label{fig:sourcedist}
\end{figure}

\section{Observations}
\label{sec:obs}
The HI absorption observations were carried out using the GMRT during
March--April 2000, and April--June 2001. On an average, we used
$\sim$20 antennas in the final analysis, though the actual number varied from
12 to 25. We used a baseband of width 2 MHz,  which translates to
$\approx$ 422 km s$^{-1}$ in velocity and a resolution of $\sim$3.3 km
s$^{-1}$. The centre of the band was set at 1420.4 MHz.
We used one of the VLA
primary flux density calibrators (3C48/3C147/3C286) for setting the flux 
density scale.
Since all the program sources were unresolved by the GMRT, they also served as
phase calibrators. Bandpass calibration was carried out once every two hours
for 10 minutes using 3C286, towards which no HI absorption was detected down to
an rms in optical depth of $\tau_{\scriptscriptstyle{\rm{HI}}}$ $\sim$0.002.
On source integration time ranged from 10 to 60 minutes, 
depending on its strength. The rms sensitivity in optical depth varied from
0.002 to 0.008 towards different sources, with a mean value $\sim$0.003.
A summary of the observational setup is given in Table \ref{tab:setup}.

\section{Data Analysis}
\label{sec:data}
The data were analysed using the Astronomical Image
Processing System (AIPS) developed by the National Radio Astronomy Observatory.
The observing band was found to be free from any kind of interference.
The resulting data set consisted of 102 image cubes. The full width at
half maximum of the synthesized beam width
was in the range $\sim$6$^{''}$ to $\sim$25$^{''}$,
depending on the number and locations of available antennas.
Continuum subtraction was carried out by fitting a second order
baseline to the line-free channels in the visibility domain and subtracting the
best fit continuum from all the channels. 
Such a second order fit to the spectral baseline can result in 
the removal of broad and shallow absorption features. From the nature of the
baselines fitted to the present dataset, we infer that any spectral features
with FWHM $\gtsim$50 km s$^{-1}$ and $\tau_{\scriptscriptstyle{\rm{HI}}}$ 
$\ltsim$0.03 would not be detected. However, the main aim of the
present survey is to study the narrow absorption lines arising from the
diffuse features in the CNM and the FWHM of such features are
usually $\ltsim$10 km s$^{-1}$. The second order fit also helps to 
achieve a better spectral dynamic range over the 2 MHz bandwidth. The
resulting spectral dynamic range was $\sim$500. However, for a few lines of
sight, the bandpass errors were much larger and only a limited
part of the observing band was found to be usable. 
To study the individual HI absorption components multiple Gaussian profiles
were fitted to the absorption line spectra using the Groningen Image
Processing System (GIPSY).

\begin{table}
\caption{The Observational Setup.}
\label{tab:setup}
\begin{center}
\begin{tabular}{ r l } \hline \hline
Telescope & GMRT \\
System temperature & $\sim$70 K \\
Aperture efficiency & $\sim$40$\%$ \\
Baseband bandwidth & 2.0 MHz \\
Number of channels & 128 \\
Velocity resolution& 3.3 km s$^{-1}$ \\
On source integration time & $\sim$10 to 60 minutes \\
rms noise (1 hr Integration time) & $\sim$2 mJy beam$^{-1}$ channel$^{-1}$ \\
\hline \hline
\end{tabular}
\end{center}
\end{table}

For an optically thin HI gas, the radiative transfer equation has a solution of
the form ( Spitzer, \nocite{ism78}1978)

\begin{equation}
\label{eqn:tbts}
T_B = T_S(1 - e^{-\tau})
\end{equation}

where $T_B$ is the brightness temperature, $T_S$ is the spin
(excitation) temperature and $\tau$ is the optical depth of the HI 21-cm line.
Hence, knowing the line brightness temperature T$_{B}$, which is obtained from
the HI emission profiles, and the optical depth $\tau$, obtained from HI
absorption, one can estimate the spin temperature T$_{S}$ of HI gas.

Apart from the new HI absorption measurements with the GMRT, we have
extracted the HI emission profiles towards these lines of sight (if available)
from the Leiden-Dwingeloo sky survey (Hartmann \& Burton, \nocite{lds}1995).
We have used Gaussian fitting to the spectra to separate the profiles into
discrete components. The HI emission
profiles were Hanning smoothed over two channels, with a resulting velocity
resolution of $\sim$2 km s$^{-1}$. 
It is well known that the HI emission 
features are broader than the corresponding absorption features. This difference
in the velocity width usually range from 0 $-$ 5 km s$^{-1}$ (Radhakrishnan et al
\nocite{rad72b}1972b). It is also 
known that the wider HI emission features (FWHM $>$ 20 km s$^{-1}$) arise in the
warm neutral medium (WNM) (Radhakrishnan et al, \nocite{rad72b}1972b).
To identify the HI emission and absorption lines that arise from the
same cloud, we have adopted the following considerations:
\begin{itemize}
\item{The central velocities of the HI emission and absorption features
are within the channel width, $\sim$3.3 km s$^{-1}$, of the GMRT observations, and} 
\item{The difference between the widths of emission and the corresponding absorption
line is less than $\sim$ 5 km s$^{-1}$}
\end{itemize}
If both the above conditions were satisfied, we assume the features to arise from
the same cloud. The second criterion excludes those instances where the velocity 
of a CNM absorption line match that of a WNM emission component. 
We have used the fitted values of $\tau$ and $T_{B}$ to
calculate the spin temperatures of HI features which are identified in both
HI emission and absorption
profiles. In those cases, where no HI absorption feature was detected from our survey
corresponding to the HI emission feature in the LDS survey data, we have
estimated the lower limit for the spin temperature of the gas.

\begin{figure}[h]
\begin{center}
\includegraphics[width=6.5cm,height=11.0cm,angle=-90]{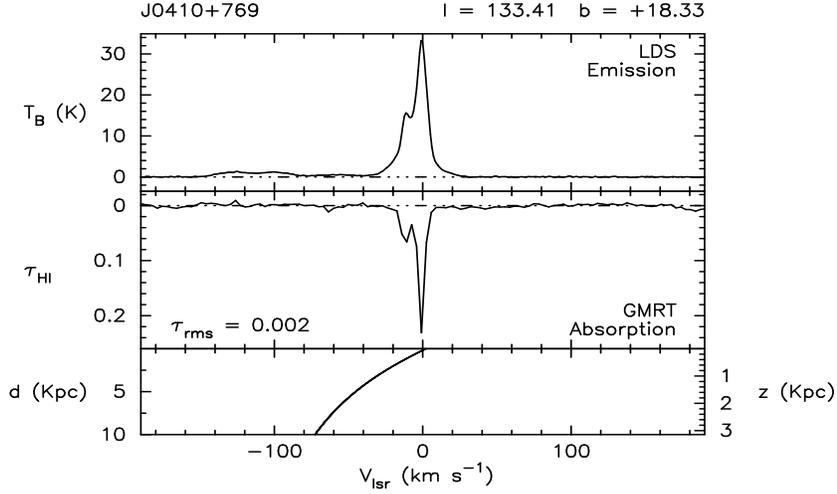}
\end{center}
\caption{The HI optical depth spectrum from the GMRT towards one of the
102 sources (middle panel) and the corresponding HI brightness temperature profile
from the Leiden-Dwingeloo survey (top panel). The lower panel is the Galactic rotation
curve for the given line of sight obtained from the Galactic rotation model (Brand
\& Blitz, 1993). The heliocentric distance as a function of radial velocity
is labeled on the left of the lower panel and the corresponding height above the mid-plane
of the Galaxy is labeled on the right side.}
\label{fig:sample}
\end{figure}

\section{The Spectra}
\label{sec:spectra}
Figure \ref{fig:sample} is a sample spectrum of HI optical depth and
the corresponding HI emission profile. In the figure, The lower panel is the 
Galactic rotation
curve for the given line of sight obtained from the Galactic rotation model by Brand
\& Blitz \nocite{bb93}(1993). We have used 
R$_{\scriptscriptstyle{\rm{0}}}$ = 8.5 Kpc as the Galacto-centric 
distance and $\Theta_{\scriptscriptstyle{\rm{0}}}$ = 220 km s$^{-1}$ as the solar
orbital velocity around the Galactic centre. The rest of the spectra are given 
in Appendix B and the
summary of the Gaussian fitting results are given in Appendix C. In all we
have obtained 126 spectral components in HI absorption from GMRT and 478 components in
HI emission from the LDS survey. The fitted parameters for the discrete
components in HI absorption and emission were used to estimate the spin
temperature of the respective features. A detailed analysis and interpretation
of these spectra are given in an accompanying paper (Mohan et al, 
\nocite{me2004b}2004; this volume).\\
~\\
\noindent{{\textbf{Acknowledgements:}}} \\
\noindent{
We wish to thank C.R. Subrahmanya for useful discussions related to
the GMRT offline software. We thank the referee,
Miller Goss, for detailed comments and constructive criticisms
resulting in an improved version of this paper. We thank the staff of the 
GMRT who made these observations possible. The GMRT is operated by the 
National Centre for Radio Astrophysics of the Tata Institute of Fundamental 
Research. This research has made use of NASA's Astrophysics Data System.}

\clearpage

\noindent \textbf {\large{Appendix A: The Source list}}
\vspace{\baselineskip} \\
\noindent
The list of sources observed with the GMRT. The sixth column gives the rms 
optical depth in the HI absorption spectra and the last column lists the 
observed flux densities of the respective sources.\\
\begin{table}[h]
\label{tab:sources}
\begin{center}
 \begin{tabular}{ l | r l l r l l | l l | l | l}                  \hline \hline
      & \multicolumn{3}{c}{$\alpha$(J2000)} & \multicolumn{3}{c|}
      {$\delta$ (J2000)}     &  l & b &
      $\tau_{\scriptscriptstyle{\rm{rms}}}$ &
      S \\
      Source & (h & m & s) & ($\circ$ & ${\prime}$ & ${\prime\prime}$)
      & ($\circ$) & ($\circ$) &       &  (Jy) \\ \hline
J0010--418 & 00 & 10 & 52.52 & --41 & 53 & 10.8 & 329.68 & --73.07 & 0.002 & 4.55 \\
J0022+002  & 00 & 22 & 25.43 &  +00 & 14 & 56.1 & 107.46 & --61.75 & 0.003 & 3.11 \\
J0024--420 & 00 & 24 & 42.99 & --42 & 02 & 04.0 & 321.35 & --74.12 & 0.003 & 2.17 \\
J0025--260 & 00 & 25 & 49.17 & --26 & 02 & 12.7 &  42.27 & --84.17 & 0.003 & 7.12 \\
J0029+349  & 00 & 29 & 14.24 &  +34 & 56 & 32.2 & 117.79 & --27.71 & 0.002 & 2.03 \\
J0059+001  & 00 & 59 & 05.51 &  +00 & 06 & 51.6 & 127.11 & --62.70 & 0.004 & 2.64 \\
J0116--208 & 01 & 16 & 51.40 & --20 & 52 & 06.8 & 167.11 & --81.47 & 0.003 & 3.91 \\
J0119+321  & 01 & 19 & 35.00 &  +32 & 10 & 50.1 & 129.83 & --30.31 & 0.003 & 3.12 \\
J0137+331  & 01 & 37 & 41.30 &  +33 & 09 & 35.1 & 133.96 & --28.72 & 0.003 &15.9 \\
J0204+152  & 02 & 04 & 50.41 &  +15 & 14 & 11.0 & 147.93 & --44.04 & 0.003 & 3.96 \\
J0204--170 & 02 & 04 & 57.67 & --17 & 01 & 19.8 & 185.99 & --70.23 & 0.003 & 1.19 \\
J0237+288  & 02 & 37 & 52.41 &  +28 & 48 & 09.0 & 149.47 & --28.53 & 0.003 & 2.10 \\
J0238+166  & 02 & 38 & 38.93 &  +16 & 36 & 59.3 & 156.77 & --39.11 & 0.004 & 1.05 \\
J0240--231 & 02 & 40 & 08.17 & --23 & 09 & 15.7 & 209.79 & --65.13 & 0.003 & 5.50 \\
J0318+164  & 03 & 18 & 57.80 &  +16 & 28 & 32.7 & 166.64 & --33.60 & 0.002 & 8.65 \\
J0321+123  & 03 & 21 & 53.10 &  +12 & 21 & 14.0 & 170.59 & --36.24 & 0.002 & 2.07 \\
J0323+055  & 03 & 23 & 20.26 &  +05 & 34 & 11.9 & 176.98 & --40.84 & 0.003 & 3.04 \\
J0329+279  & 03 & 29 & 57.67 &  +27 & 56 & 15.5 & 160.70 & --23.07 & 0.003 & 1.40 \\
J0336+323  & 03 & 36 & 30.11 &  +32 & 18 & 29.3 & 159.00 & --18.77 & 0.004 & 2.71 \\
J0348+338  & 03 & 48 & 46.90 &  +33 & 53 & 15.0 & 160.04 & --15.91 & 0.003 & 2.32 \\
J0403+260  & 04 & 03 & 05.59 &  +26 & 00 & 01.5 & 168.03 & --19.65 & 0.004 & 1.08 \\
J0409+122  & 04 & 09 & 22.01 &  +12 & 17 & 39.8 & 180.12 & --27.90 & 0.003 & 1.46 \\
J0410+769  & 04 & 10 & 45.61 &  +76 & 56 & 45.3 & 133.41 &  +18.33 & 0.002 & 5.70 \\
J0424+020  & 04 & 24 & 08.56 &  +02 & 04 & 24.9 & 192.04 & --31.10 & 0.004 & 1.33 \\
J0431+206  & 04 & 31 & 03.76 &  +20 & 37 & 34.3 & 176.81 & --18.56 & 0.003 & 3.40 \\
J0440--435 & 04 & 40 & 17.18 & --43 & 33 & 08.6 & 248.41 & --41.57 & 0.003 & 3.28 \\
J0453--281 & 04 & 53 & 14.65 & --28 & 07 & 37.3 & 229.09 & --37.02 & 0.004 & 2.18 \\
J0459+024  & 04 & 59 & 52.05 &  +02 & 29 & 31.2 & 197.01 & --23.34 & 0.003 & 1.93 \\
J0503+020  & 05 &  3 & 21.20 &  +02 & 03 & 04.7 & 197.91 & --22.82 & 0.004 & 2.26 \\
J0538--440 & 05 & 38 & 50.36 & --44 & 05 & 08.9 & 250.08 & --31.09 & 0.002 & 2.75 \\
J0541--056 & 05 & 41 & 38.08 & --05 & 41 & 49.4 & 210.05 & --18.11 & 0.004 & 1.42 \\
J0609--157 & 06 & 09 & 40.95 & --15 & 42 & 40.7 & 222.61 & --16.18 & 0.007 & 2.78 \\
J0614+607  & 06 & 14 & 23.87 &  +60 & 46 & 21.8 & 153.60 &  +19.15 & 0.005 & 1.17 \\
\hline \hline
\end{tabular}
\end{center}
\end{table}

\begin{table}
\begin{center}
 \begin{tabular}{ l | r l l r l l | l l | l | l}                  \hline \hline
      & \multicolumn{3}{c}{$\alpha$(J2000)} & \multicolumn{3}{c|}
      {$\delta$ (J2000)}     &  l & b &
      $\tau_{\scriptscriptstyle{\rm{rms}}}$ &
      S \\
      Source & (h & m & s) & ($\circ$ & ${\prime}$ & ${\prime\prime}$)
      & ($\circ$) & ($\circ$) &       &  (Jy) \\ \hline
J0713+438  & 07 & 13 & 38.16 &  +43 & 49 & 17.2 & 173.79 &  +22.20 & 0.004 & 2.33 \\
J0814+459  & 08 & 14 & 30.31 &  +45 & 56 & 39.5 & 173.90 &  +33.17 & 0.005 & 1.13 \\ 
J0825+031  & 08 & 25 & 50.34 &  +03 & 09 & 24.5 & 221.22 &  +22.39 & 0.002 & 1.12 \\
J0834+555  & 08 & 34 & 54.90 &  +55 & 34 & 21.1 & 162.23 &  +36.56 & 0.002 & 9.15 \\
J0842+185  & 08 & 42 & 05.09 &  +18 & 35 & 41.0 & 207.28 &  +32.48 & 0.004 & 1.17 \\
J0854+201  & 08 & 54 & 48.87 &  +20 & 06 & 30.6 & 206.81 &  +35.82 & 0.002 & 1.65 \\
J0921--263 & 09 & 21 & 29.35 & --26 & 18 & 43.4 & 255.07 &  +16.48 & 0.003 & 1.41 \\
J0958+324  & 09 & 58 & 20.95 &  +32 & 24 & 02.2 & 194.17 &  +52.32 & 0.005 & 1.47 \\
J1018--317 & 10 & 18 & 09.28 & --31 & 44 & 14.1 & 268.61 &  +20.73 & 0.002 & 3.44 \\
J1057--245 & 10 & 57 & 55.42 & --24 & 33 & 48.9 & 272.47 &  +31.51 & 0.004 & 1.10 \\
J1111+199  & 11 & 11 & 20.07 &  +19 & 55 & 36.0 & 225.01 &  +66.00 & 0.003 & 1.52 \\
J1119--030 & 11 & 19 & 25.30 & --03 & 02 & 51.3 & 263.01 &  +52.54 & 0.002 & 1.44 \\
J1120--251 & 11 & 20 & 09.12 & --25 & 08 & 07.6 & 278.09 &  +33.30 & 0.003 & 1.31 \\
J1125+261  & 11 & 25 & 53.71 &  +26 & 10 & 20.0 & 210.92 &  +70.89 & 0.003 & 0.96 \\
J1130--148 & 11 & 30 & 07.05 & --14 & 49 & 27.4 & 275.28 &  +43.64 & 0.002 & 5.96 \\
J1146+399  & 11 & 46 & 58.30 &  +39 & 58 & 34.3 & 164.95 &  +71.47 & 0.009 & 0.47 \\
J1154--350 & 11 & 54 & 21.79 & --35 & 05 & 29.0 & 289.93 &  +26.34 & 0.002 & 5.48 \\
J1221+282  & 12 & 21 & 31.69 &  +28 & 13 & 58.5 & 201.74 &  +83.29 & 0.004 & 1.03 \\
J1235--418 & 12 & 35 & 41.93 & --41 & 53 & 18.0 & 299.80 &  +20.89 & 0.003 & 1.57 \\
J1254+116  & 12 & 54 & 38.26 &  +11 & 41 & 05.9 & 305.87 &  +74.54 & 0.005 & 0.93 \\
J1257--319 & 12 & 57 & 59.06 & --31 & 55 & 16.9 & 304.55 &  +30.93 & 0.003 & 1.21 \\
J1316--336 & 13 & 16 & 07.99 & --33 & 38 & 59.2 & 308.80 &  +28.94 & 0.006 & 1.11 \\
J1344+141  & 13 & 44 & 23.74 &  +14 & 09 & 14.9 & 349.16 &  +72.09 & 0.003 & 1.34 \\
J1351--148 & 13 & 51 & 52.65 & --14 & 49 & 14.9 & 324.03 &  +45.56 & 0.003 & 1.16 \\
J1357--154 & 13 & 57 & 11.24 & --15 & 27 & 28.8 & 325.42 &  +44.52 & 0.015 & 1.11 \\
J1435+760  & 14 & 35 & 47.10 &  +76 & 05 & 25.8 & 115.07 &  +39.40 & 0.003 & 1.19 \\
J1445+099  & 14 & 45 & 16.47 &  +09 & 58 & 36.1 &   5.79 &  +58.17 & 0.002 & 2.62 \\
J1448--163 & 14 & 48 & 15.05 & --16 & 20 & 24.5 & 339.45 &  +38.11 & 0.003 & 1.63 \\
J1506+375  & 15 & 06 & 09.53 &  +37 & 30 & 51.1 &  61.65 &  +59.90 & 0.003 & 1.15 \\
J1513+236  & 15 & 13 & 40.19 &  +23 & 38 & 35.2 &  34.77 &  +57.79 & 0.002 & 1.73 \\
J1517--243 & 15 & 17 & 41.81 & --24 & 22 & 19.5 & 340.68 &  +27.58 & 0.004 & 2.53 \\
J1520+202  & 15 & 20 & 05.49 &  +20 & 16 & 05.6 &  29.64 &  +55.42 & 0.003 & 3.17 \\
J1526--138 & 15 & 26 & 59.44 & --13 & 51 & 00.1 & 350.48 &  +34.29 & 0.002 & 2.63 \\
J1553+129  & 15 & 53 & 32.70 &  +12 & 56 & 51.7 &  23.79 &  +45.22 & 0.003 & 1.49 \\    
J1554--270 & 15 & 54 & 02.49 & --27 & 04 & 40.2 & 345.68 &  +20.27 & 0.003 & 1.52 \\
J1557--000 & 15 & 57 & 51.43 &  +00 & 01 & 50.4 &   9.58 &  +37.68 & 0.004 & 0.95 \\
J1602+334  & 16 & 02 & 07.26 &  +33 & 26 & 53.1 &  53.73 &  +48.71 & 0.002 & 3.33 \\
J1609+266  & 16 & 09 & 13.32 &  +26 & 41 & 29.0 &  44.17 &  +46.20 & 0.003 & 4.57 \\
J1613+342  & 16 & 13 & 41.06 &  +34 & 12 & 47.9 &  55.15 &  +46.38 & 0.001 & 5.35 \\
\hline \hline
\end{tabular}
\end{center}
\end{table}

\begin{table}
\begin{center}
 \begin{tabular}{ l | r l l r l l | l l | l | l}                  \hline \hline
      & \multicolumn{3}{c}{$\alpha$(J2000)} & \multicolumn{3}{c|}
      {$\delta$ (J2000)}     &  l & b &
      $\tau_{\scriptscriptstyle{\rm{rms}}}$ &
      S \\
      Source & (h & m & s) & ($\circ$ & ${\prime}$ & ${\prime\prime}$)
      & ($\circ$) & ($\circ$) &       &  (Jy) \\ \hline
J1634+627  & 16 & 34 & 33.80 &  +62 & 45 & 35.9 &  93.61 &  +39.38 & 0.002 & 4.67 \\
J1635+381  & 16 & 35 & 15.49 &  +38 & 08 & 04.5 &  61.09 &  +42.34 & 0.002 & 3.45 \\
J1638+625  & 16 & 38 & 28.21 &  +62 & 34 & 44.3 &  93.22 &  +39.01 & 0.003 & 4.65 \\
J1640+123  & 16 & 40 & 47.93 &  +12 & 20 & 02.1 &  29.43 &  +34.51 & 0.003 & 1.87 \\
J1737+063  & 17 & 37 & 13.73 &  +06 & 21 & 03.5 &  30.15 &  +19.38 & 0.005 & 0.87 \\
J1745+173  & 17 & 45 & 35.21 &  +17 & 20 & 01.4 &  41.74 &  +22.12 & 0.004 & 1.19 \\
J1751+096  & 17 & 51 & 32.82 &  +09 & 39 & 00.7 &  34.92 &  +17.65 & 0.002 & 1.76 \\
J1800+784  & 18 & 00 & 45.68 &  +78 & 28 & 04.1 & 110.04 &  +29.07 & 0.002 & 2.71 \\   
J1845+401  & 18 & 45 & 11.12 &  +40 & 07 & 51.5 &  69.36 &  +18.21 & 0.003 & 1.41 \\
J1923--210 & 19 & 23 & 32.19 & --21 & 04 & 33.3 &  17.18 & --16.25 & 0.003 & 2.82 \\
J2005+778  & 20 & 05 & 31.00 &  +77 & 52 & 43.2 & 110.46 &  +22.73 & 0.003 & 1.38 \\
J2009+724  & 20 & 09 & 52.30 &  +72 & 29 & 19.4 & 105.36 &  +20.18 & 0.003 & 1.00 \\
J2011--067 & 20 & 11 & 14.22 & --06 & 44 & 03.6 &  36.01 & --20.80 & 0.002 & 3.39 \\
J2047--026 & 20 & 47 & 10.35 & --02 & 36 & 22.2 &  44.56 & --26.80 & 0.002 & 2.94 \\
J2130+050  & 21 & 30 & 32.88 &  +05 & 02 & 17.5 &  58.65 & --31.81 & 0.002 & 5.05 \\
J2136+006  & 21 & 36 & 38.59 &  +00 & 41 & 54.2 &  55.47 & --35.58 & 0.002 & 5.22 \\
J2137--207 & 21 & 37 & 50.00 & --20 & 42 & 31.8 &  30.35 & --45.56 & 0.002 & 3.87 \\
J2148+069  & 21 & 48 & 05.46 &  +06 & 57 & 38.6 &  63.66 & --34.07 & 0.002 & 3.74 \\
J2212+018  & 22 & 12 & 37.98 &  +01 & 52 & 51.2 &  63.68 & --42.02 & 0.001 & 3.95 \\
J2214--385 & 22 & 14 & 38.57 & --38 & 35 & 45.0 &   3.47 & --55.44 & 0.002 & 1.76 \\
J2219--279 & 22 & 19 & 40.94 & --27 & 56 & 26.9 &  22.57 & --56.48 & 0.003 & 2.55 \\
J2225--049 & 22 & 25 & 47.26 & --04 & 57 & 01.4 &  58.96 & --48.84 & 0.003 & 5.32 \\
J2232+117  & 22 & 32 & 36.41 &  +11 & 43 & 50.9 &  77.44 & --38.58 & 0.003 & 7.43 \\
J2236+284  & 22 & 36 & 22.47 &  +28 & 28 & 57.4 &  90.12 & --25.65 & 0.004 & 1.38 \\
J2246--121 & 22 & 46 & 18.23 & --12 & 06 & 51.3 &  53.87 & --57.07 & 0.003 & 2.26 \\
J2250+143  & 22 & 50 & 25.54 &  +14 & 19 & 50.6 &  83.89 & --39.20 & 0.002 & 2.16 \\
J2251+188  & 22 & 51 & 34.74 &  +18 & 48 & 40.1 &  87.35 & --35.65 & 0.002 & 3.24 \\
J2302--373 & 23 & 02 & 23.89 & --37 & 18 & 06.8 &   2.16 & --64.91 & 0.002 & 3.09 \\
J2340+135  & 23 & 40 & 33.22 &  +13 & 33 & 00.9 &  97.80 & --45.83 & 0.002 & 2.82 \\
J2341--351 & 23 & 41 & 45.89 & --35 & 06 & 22.1 &   0.45 & --73.12 & 0.004 & 2.06 \\
  \hline \hline
 \end{tabular}
\end{center}
\end{table}

\clearpage

\noindent \textbf {\large{Appendix B: The Spectra}}
\vspace{\baselineskip} \\
In this Appendix, we present the HI optical depth profiles obtained from
the high latitude Galactic HI absorption survey using the GMRT
along with the HI emission profiles in the respective lines of sight from
the Leiden-Dwingeloo survey of Galactic Neutral hydrogen (if available).
The figures are
arranged in order of increasing Right Ascension. The spectra are labeled by
the radio continuum source name in J2000.0 co-ordinates (top left) and
its Galactic co-ordinates (top right). \\

For each of the figures, the HI emission profile from the Leiden-Dwingeloo Survey is
shown in the top panel and the HI optical depth profile from the GMRT is shown in
the middle panel. The lower panel is the Galactic rotation curve for the given line of sight.
The heliocentric distance as a function of radial velocity (V$_{lsr}$) is labeled 
on the left of this
panel and the corresponding height above the mid-plane of the Galaxy is labeled on its
right side. For a few lines of sight, only the reliable part of the observing band is shown.\\

\vspace{0.5cm}
{\centering\includegraphics[height=11.33cm,width=12cm]{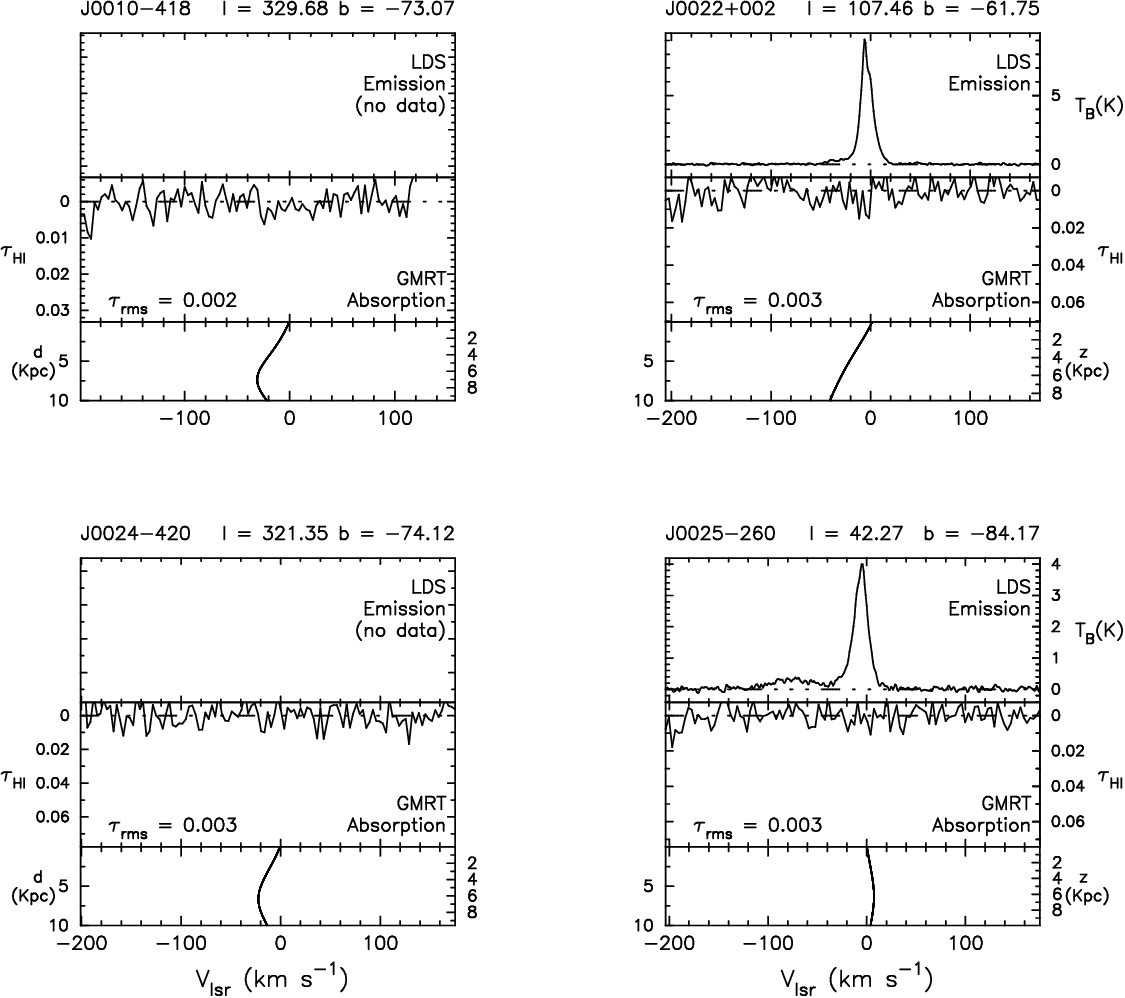}}

\clearpage

\begin{figure}
\vspace{1.0cm}
{\centering\includegraphics[height=18.5cm,width=12cm]{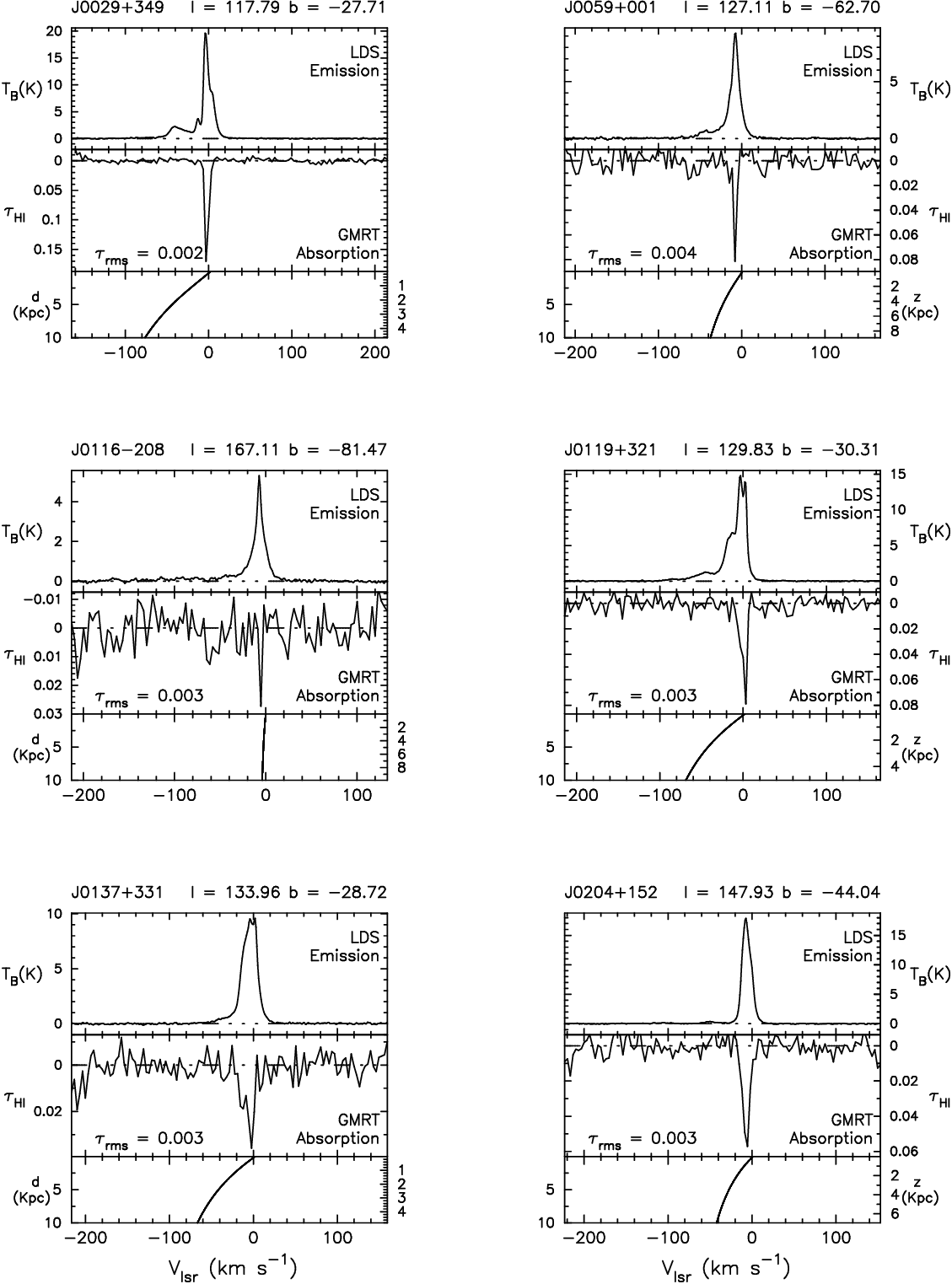}}
\end{figure}

\clearpage

\begin{figure}
\vspace{1.0cm}
{\centering\includegraphics[height=18.5cm,width=12cm]{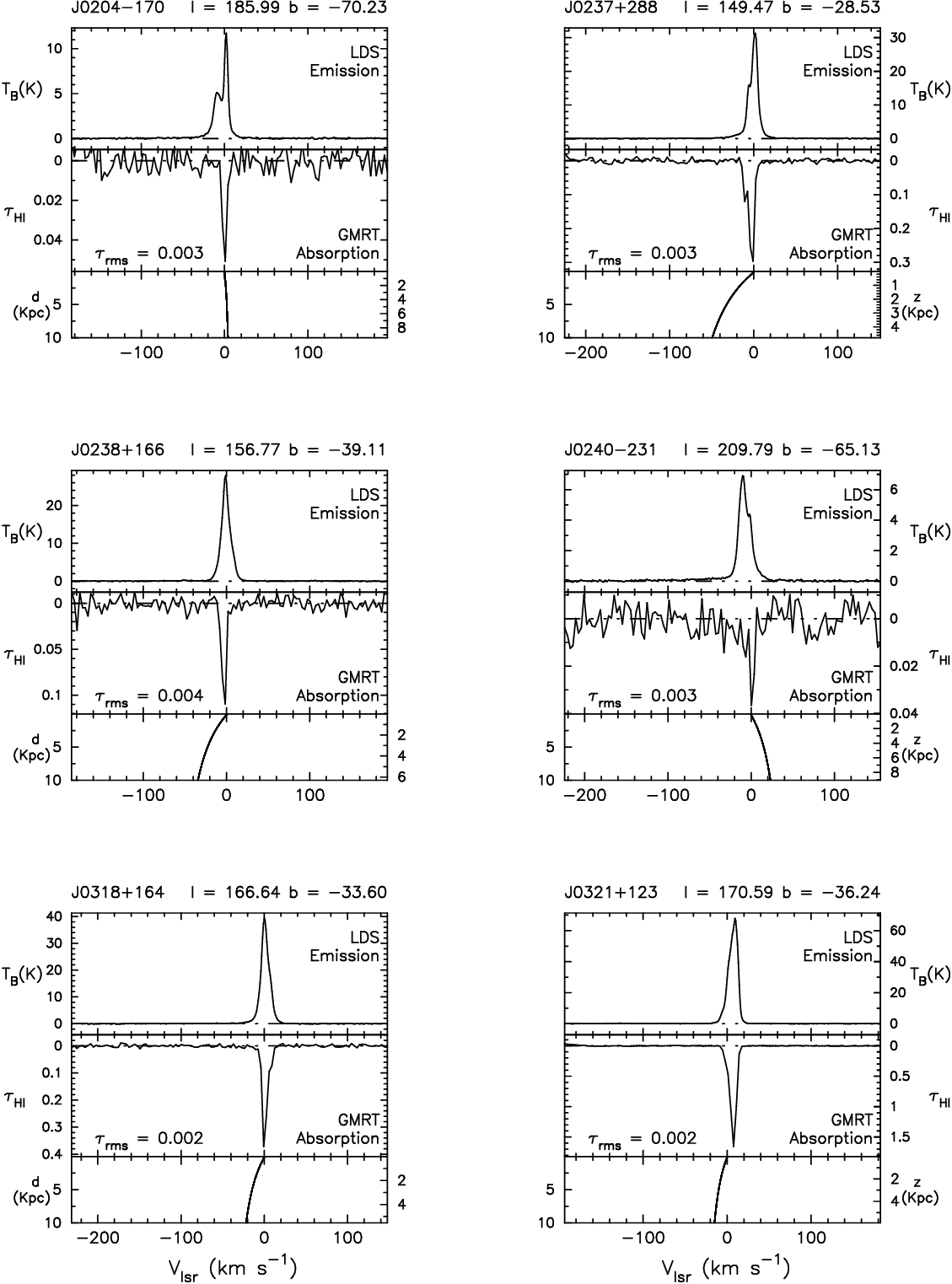}}
\end{figure}

\clearpage

\begin{figure}
\vspace{1.0cm}
{\centering\includegraphics[height=18.5cm,width=12cm]{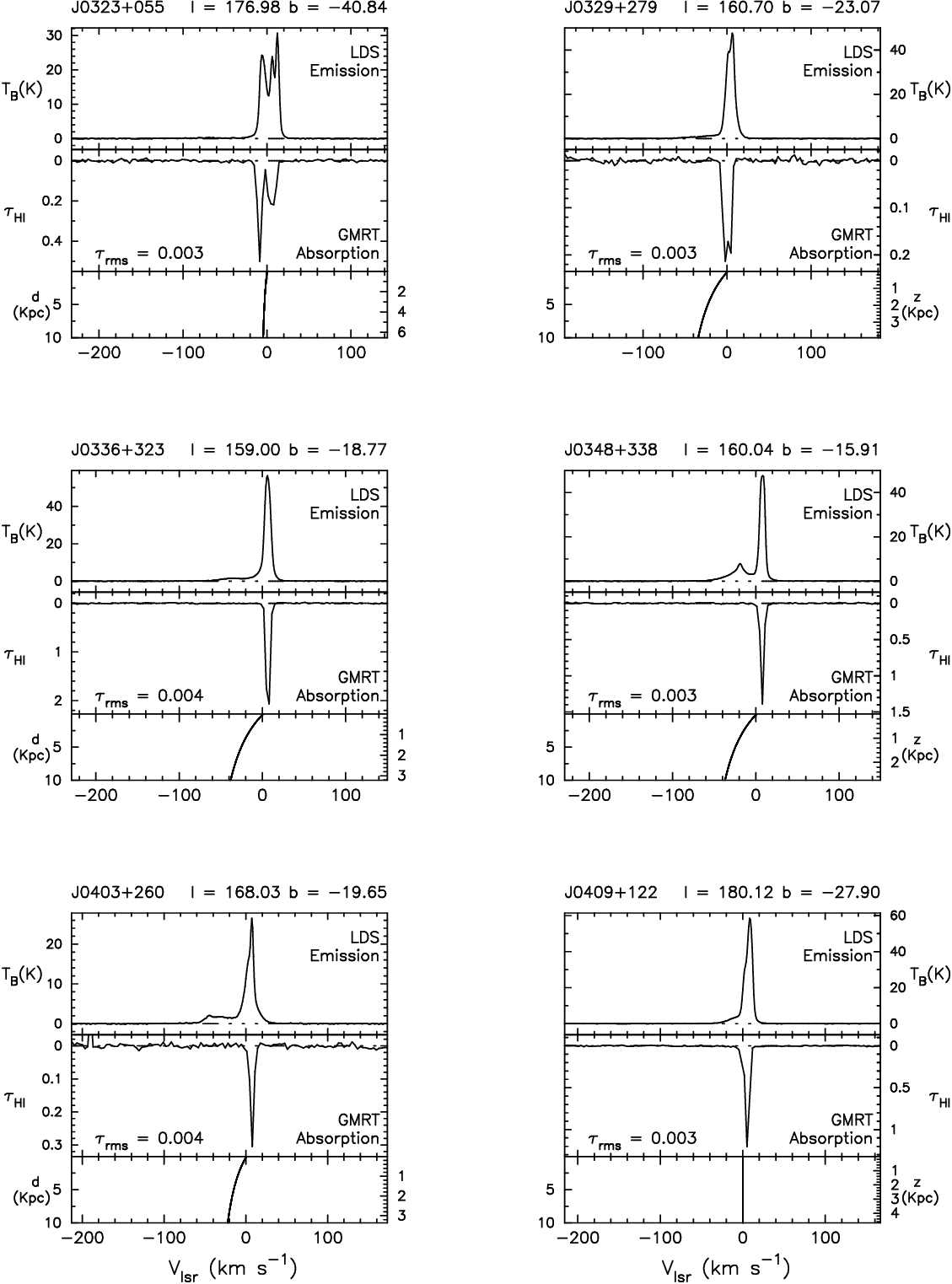}}
\end{figure}

\clearpage

\begin{figure}
\vspace{1.0cm}
{\centering\includegraphics[height=18.5cm,width=12cm]{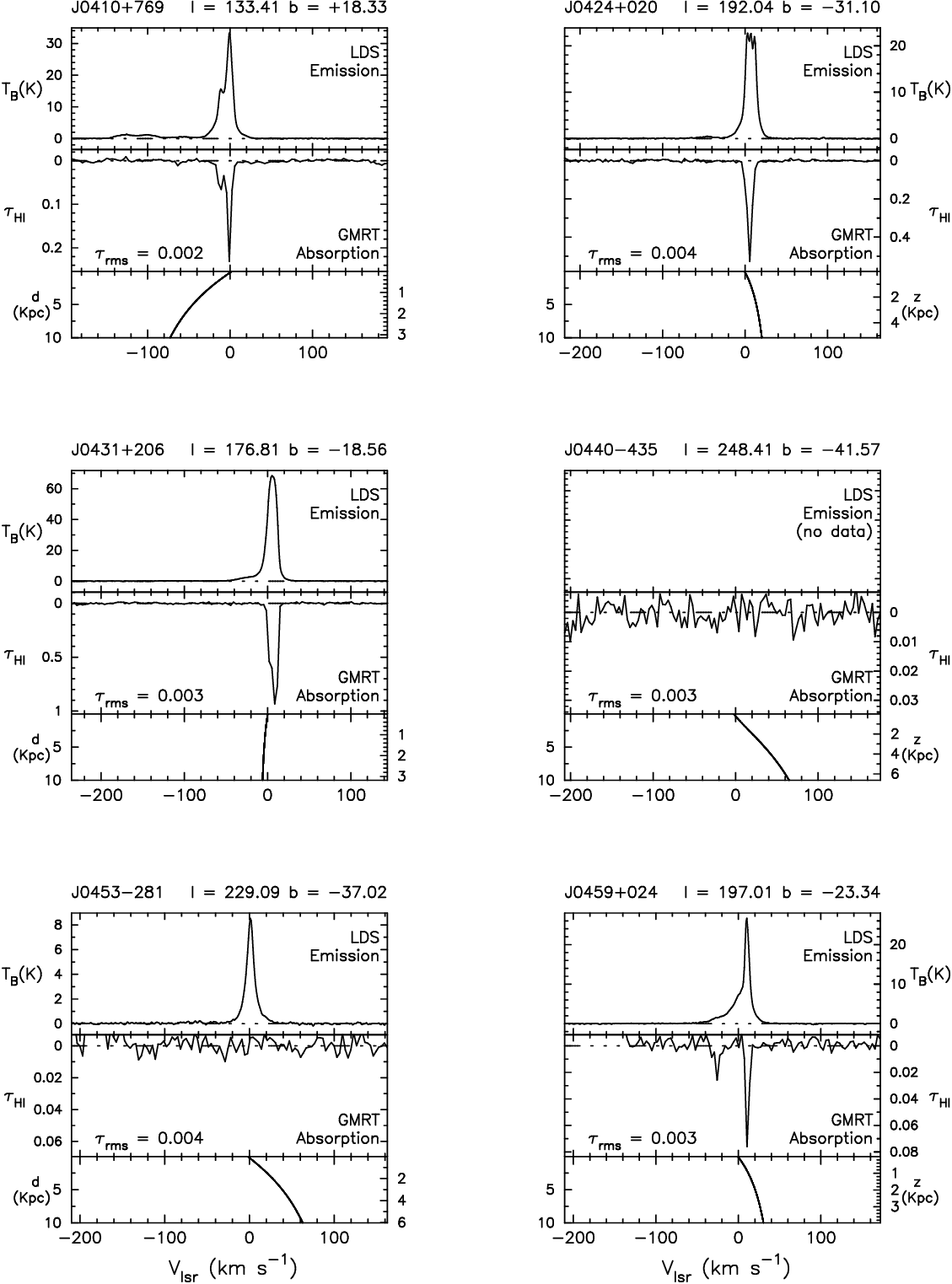}}
\end{figure}

\clearpage

\begin{figure}
\vspace{1.0cm}
{\centering\includegraphics[height=18.5cm,width=12cm]{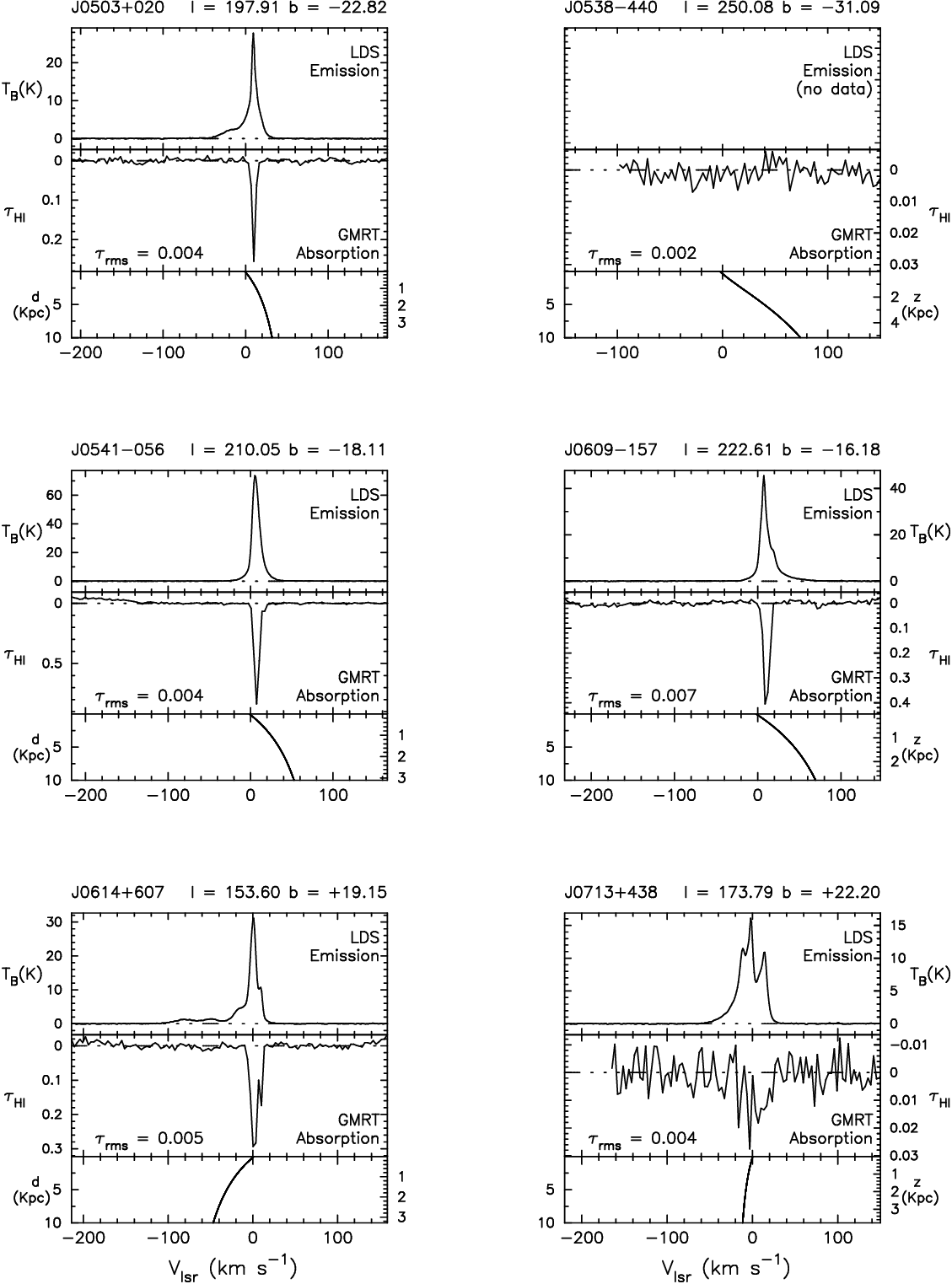}}
\end{figure}

\clearpage

\begin{figure}
\vspace{1.0cm}
{\centering\includegraphics[height=18.5cm,width=12cm]{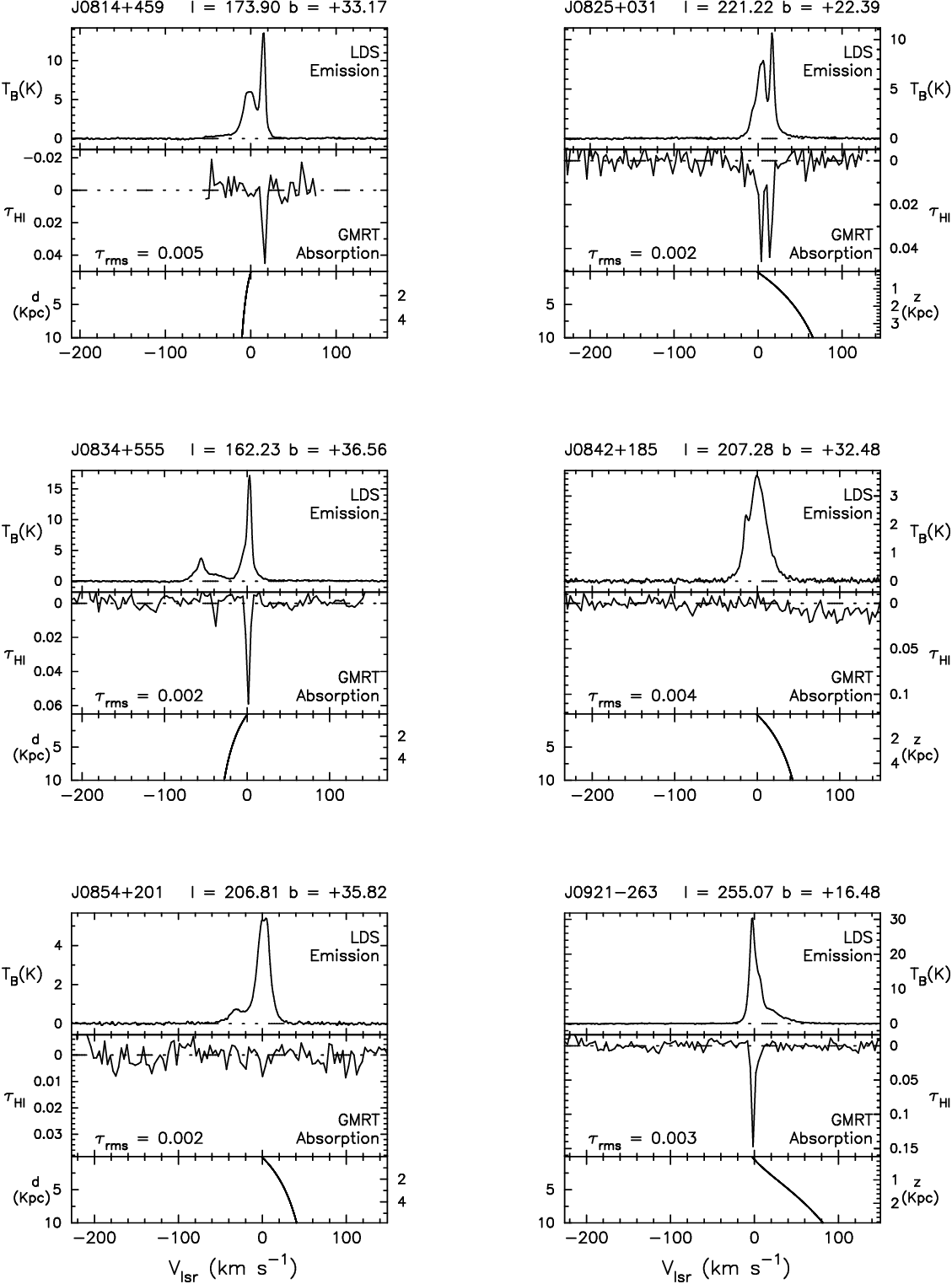}}
\end{figure}

\clearpage

\begin{figure}
\vspace{1.0cm}
{\centering\includegraphics[height=18.5cm,width=12cm]{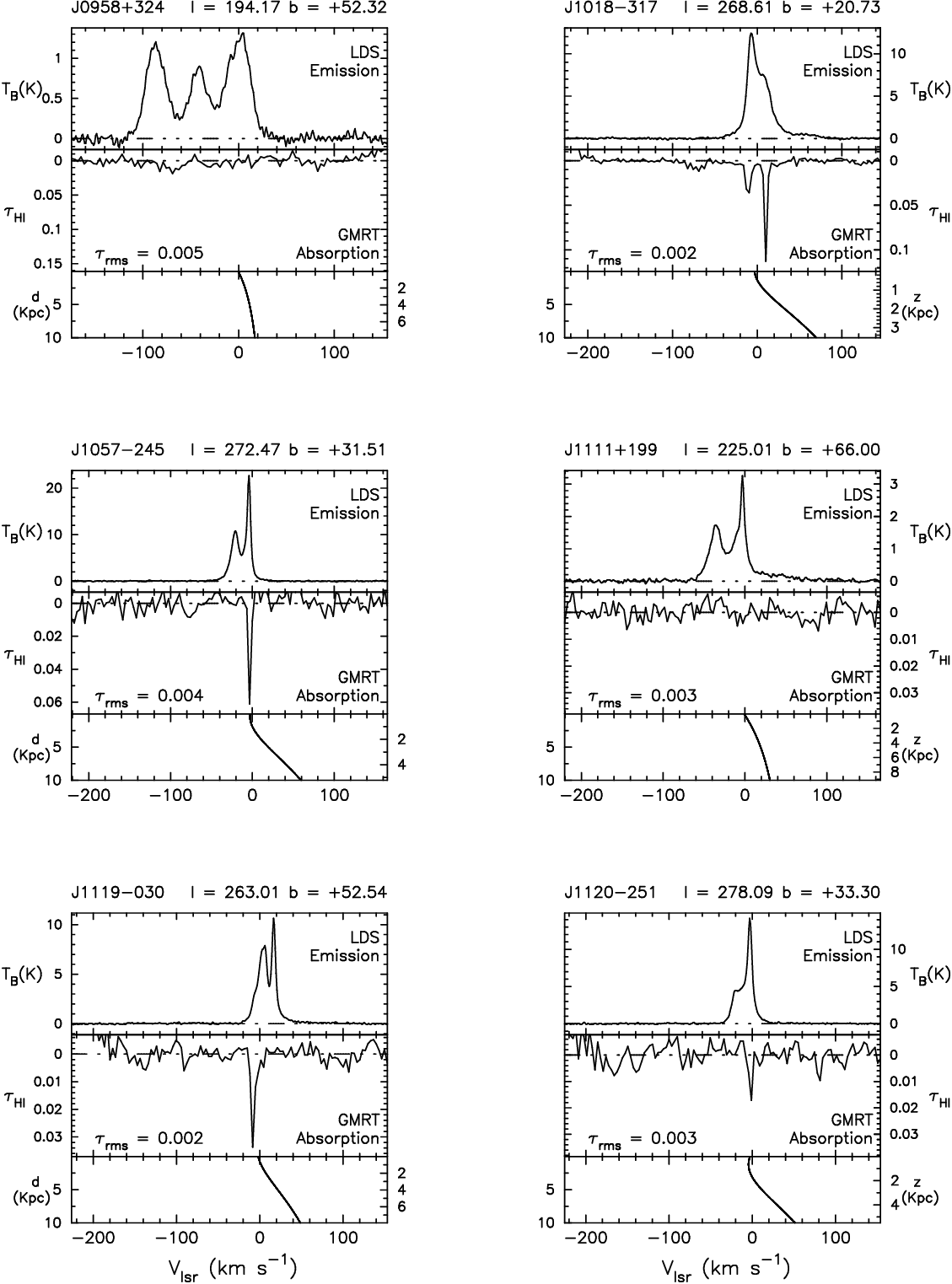}}
\end{figure}

\clearpage

\begin{figure}
\vspace{1.0cm}
{\centering\includegraphics[height=18.5cm,width=12cm]{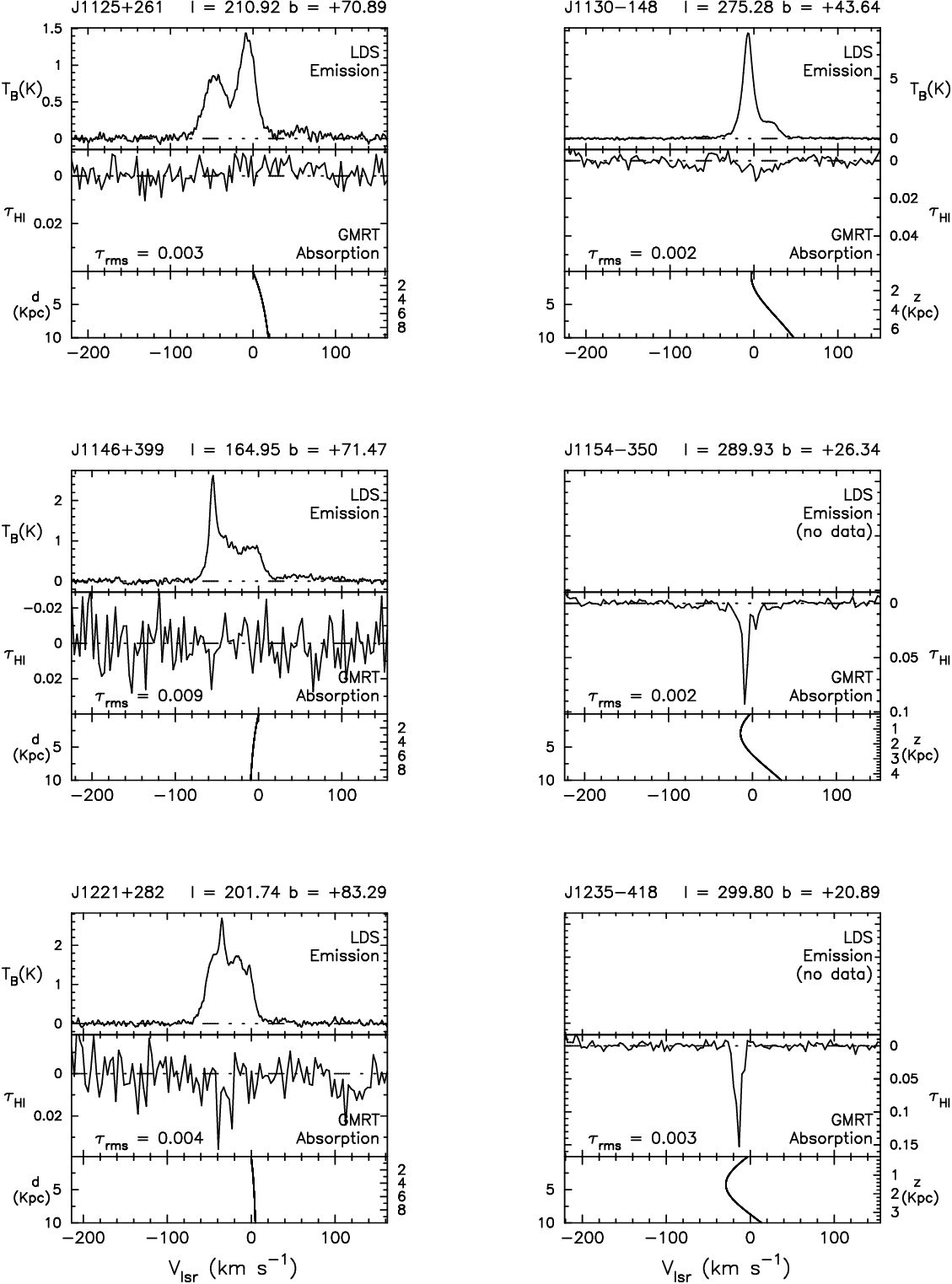}}
\end{figure}

\clearpage

\begin{figure}
\vspace{1.0cm}
{\centering\includegraphics[height=18.5cm,width=12cm]{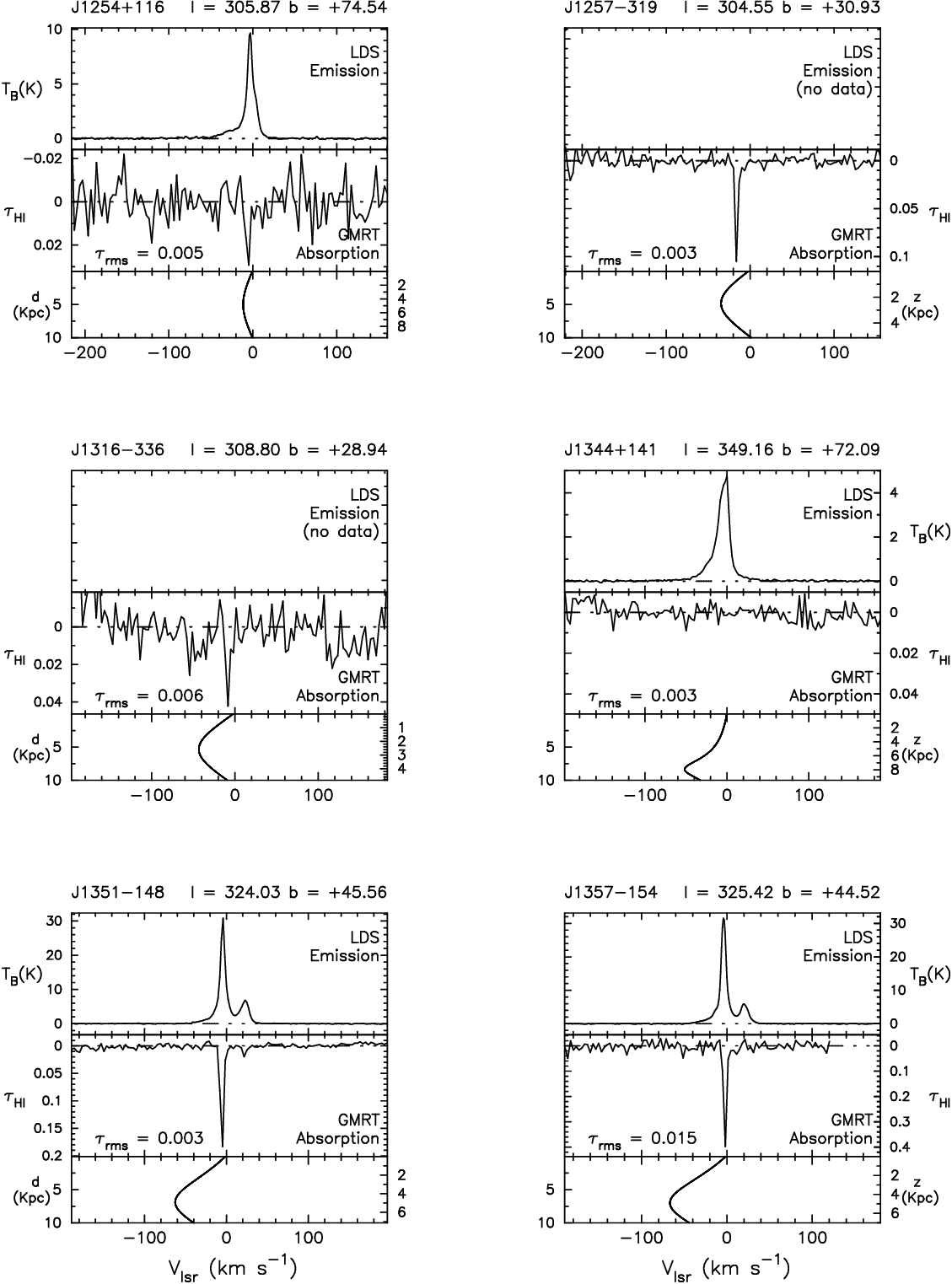}}
\end{figure}

\clearpage

\begin{figure}
\vspace{1.0cm}
{\centering\includegraphics[height=18.5cm,width=12cm]{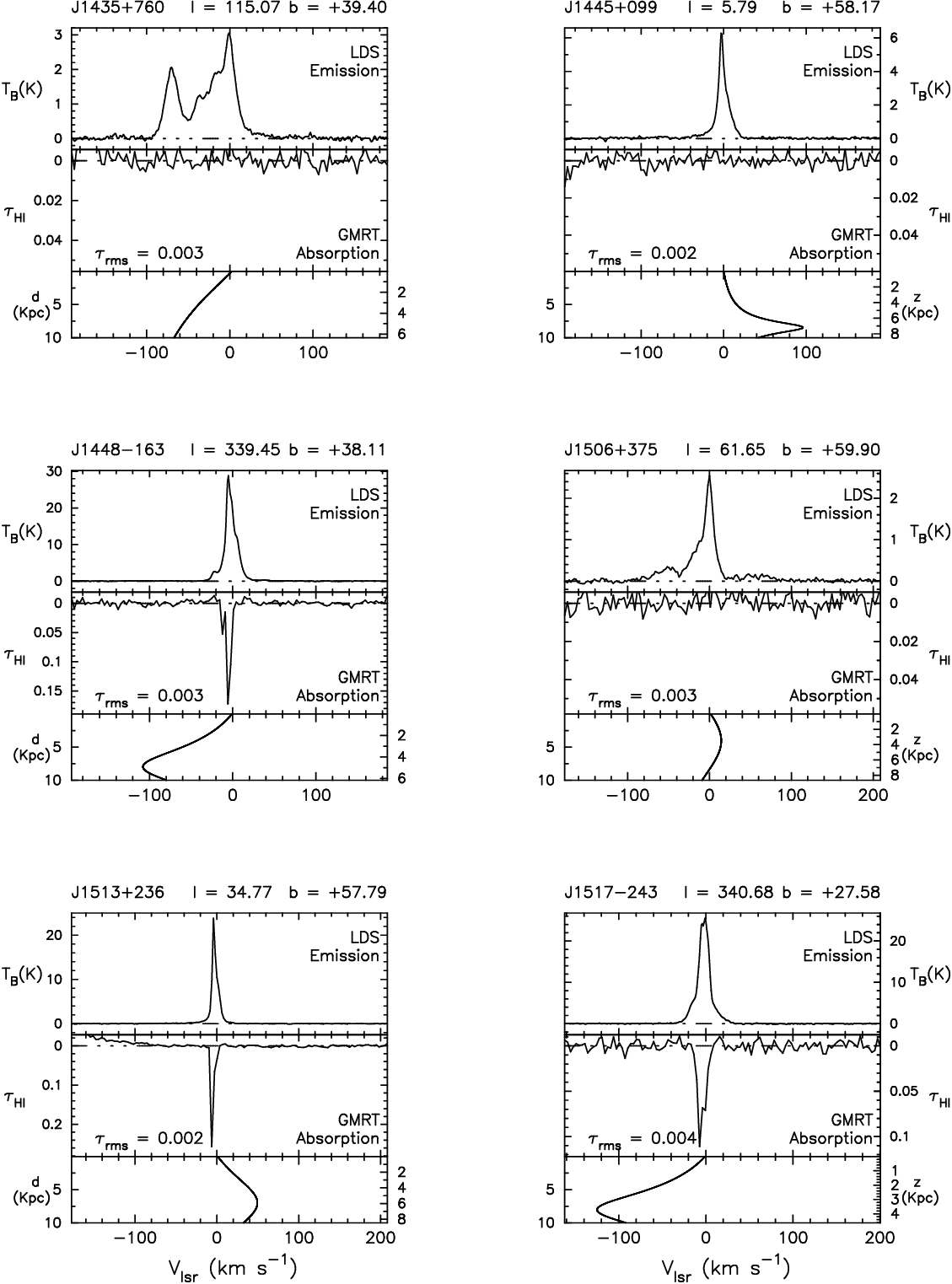}}
\end{figure}

\clearpage

\begin{figure}
\vspace{1.0cm}
{\centering\includegraphics[height=18.5cm,width=12cm]{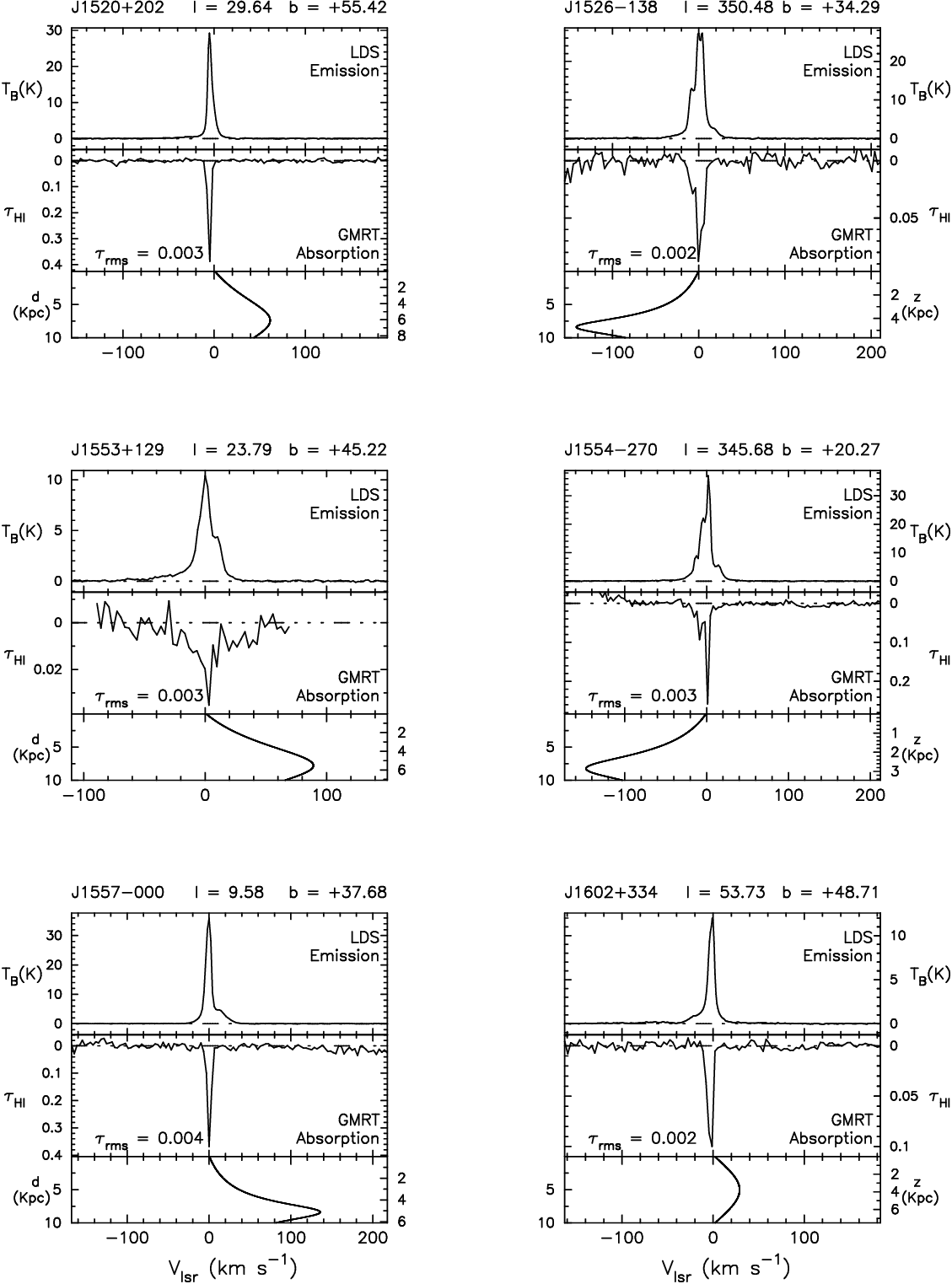}}
\end{figure}

\clearpage

\begin{figure}
\vspace{1.0cm}
{\centering\includegraphics[height=18.5cm,width=12cm]{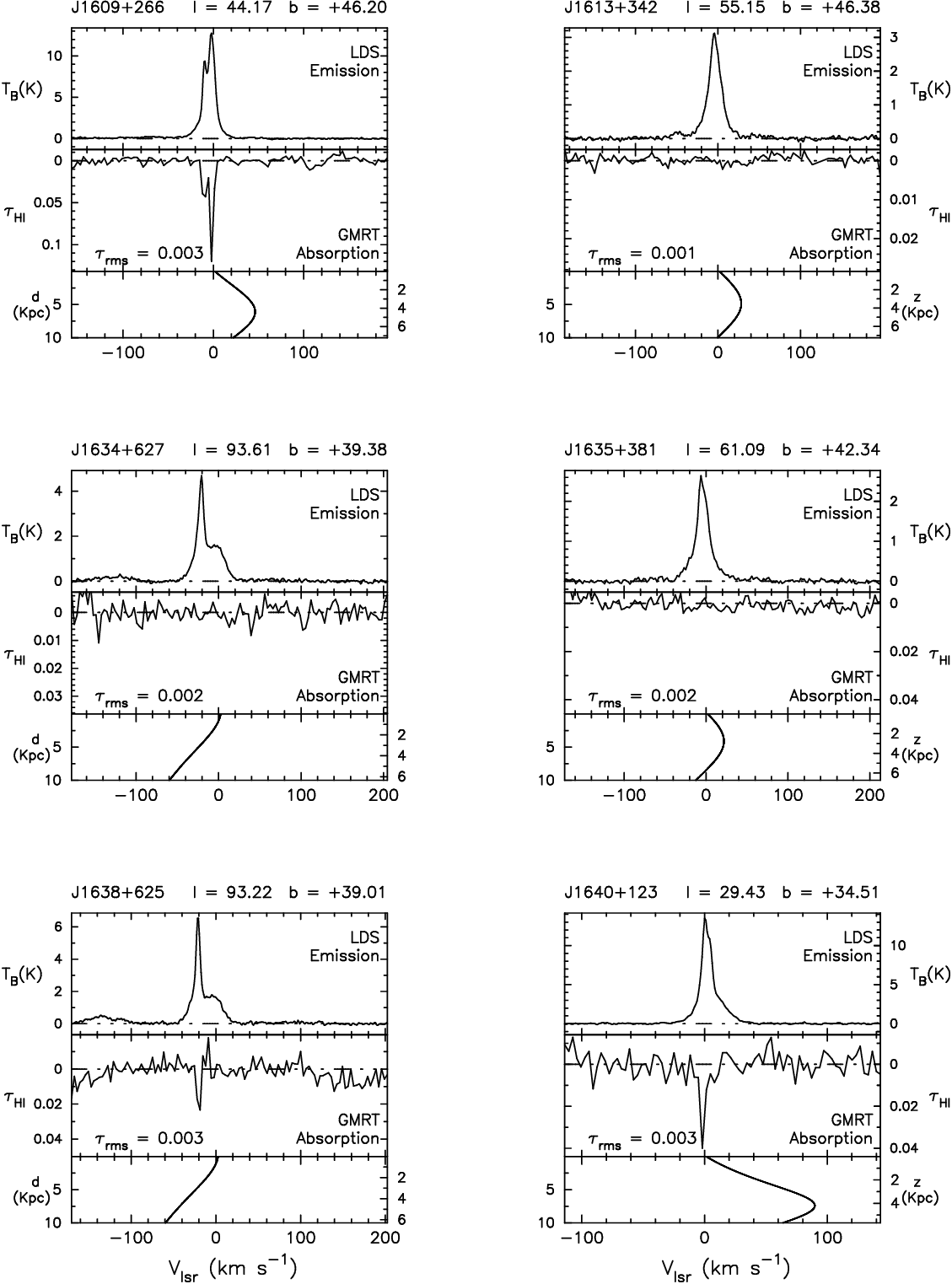}}
\end{figure}

\clearpage

\begin{figure}
\vspace{1.0cm}
{\centering\includegraphics[height=18.5cm,width=12cm]{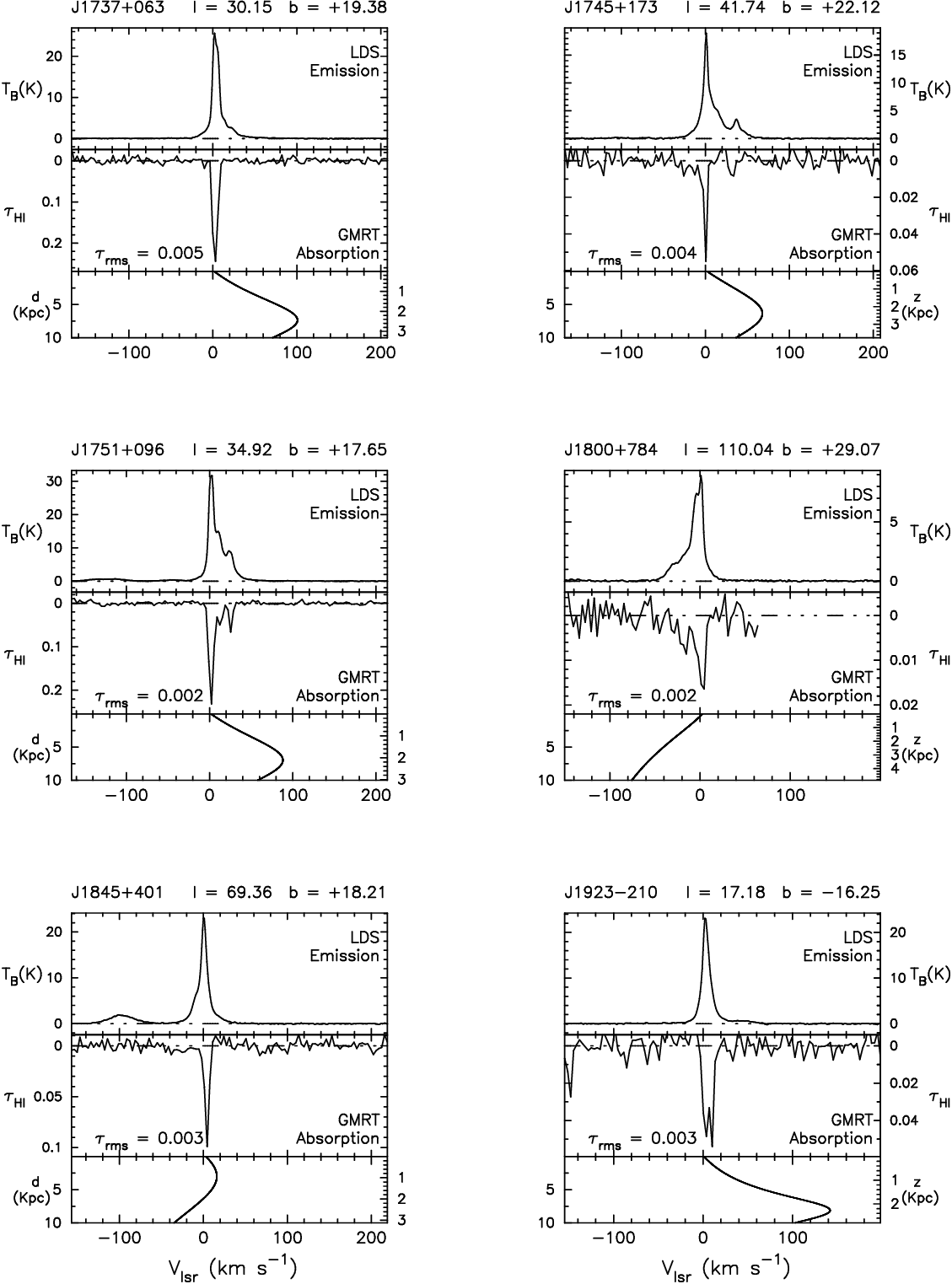}}
\end{figure}

\clearpage

\begin{figure}
\vspace{1.0cm}
{\centering\includegraphics[height=18.5cm,width=12cm]{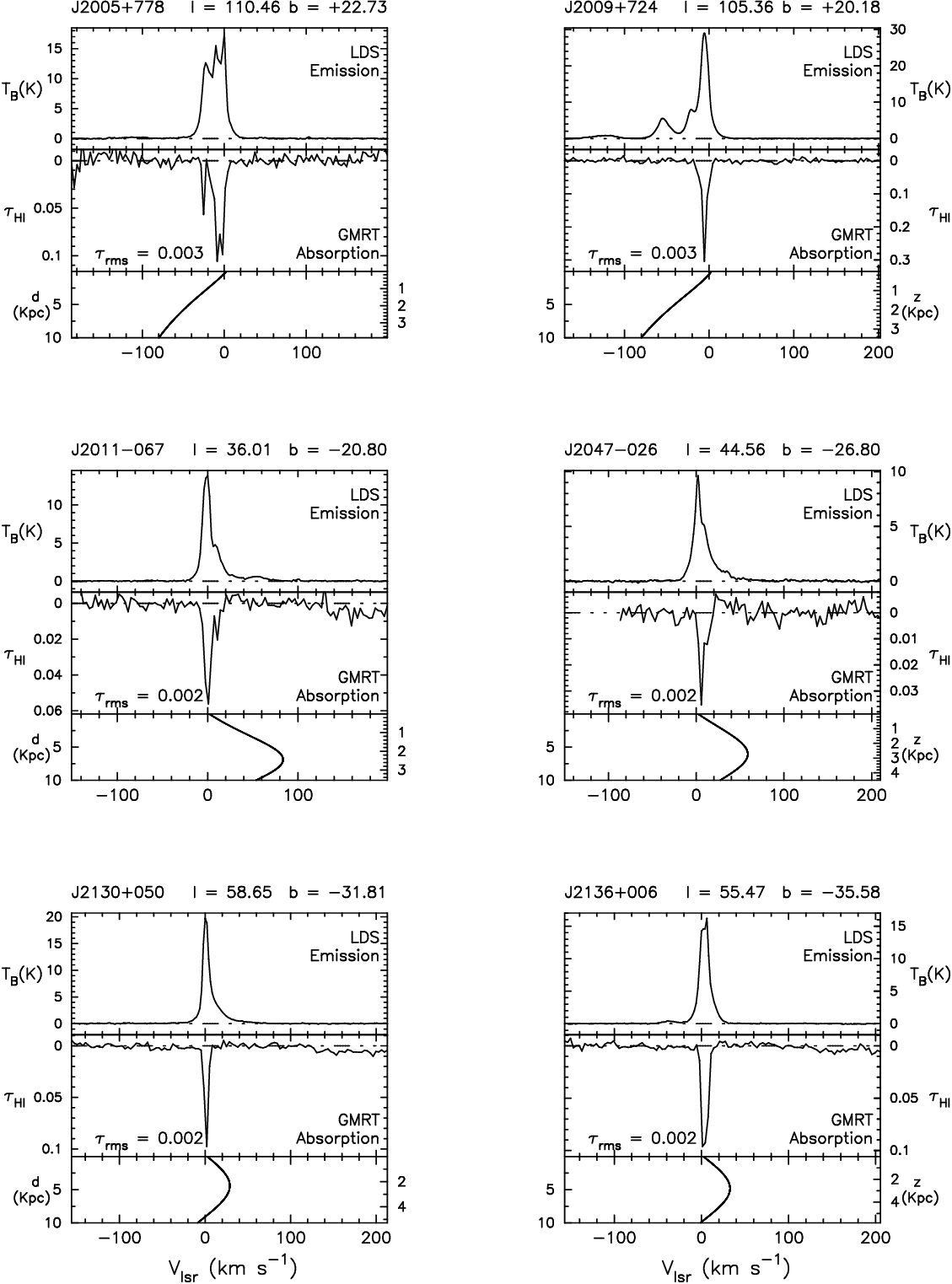}}
\end{figure}

\clearpage

\begin{figure}
\vspace{1.0cm}
{\centering\includegraphics[height=18.5cm,width=12cm]{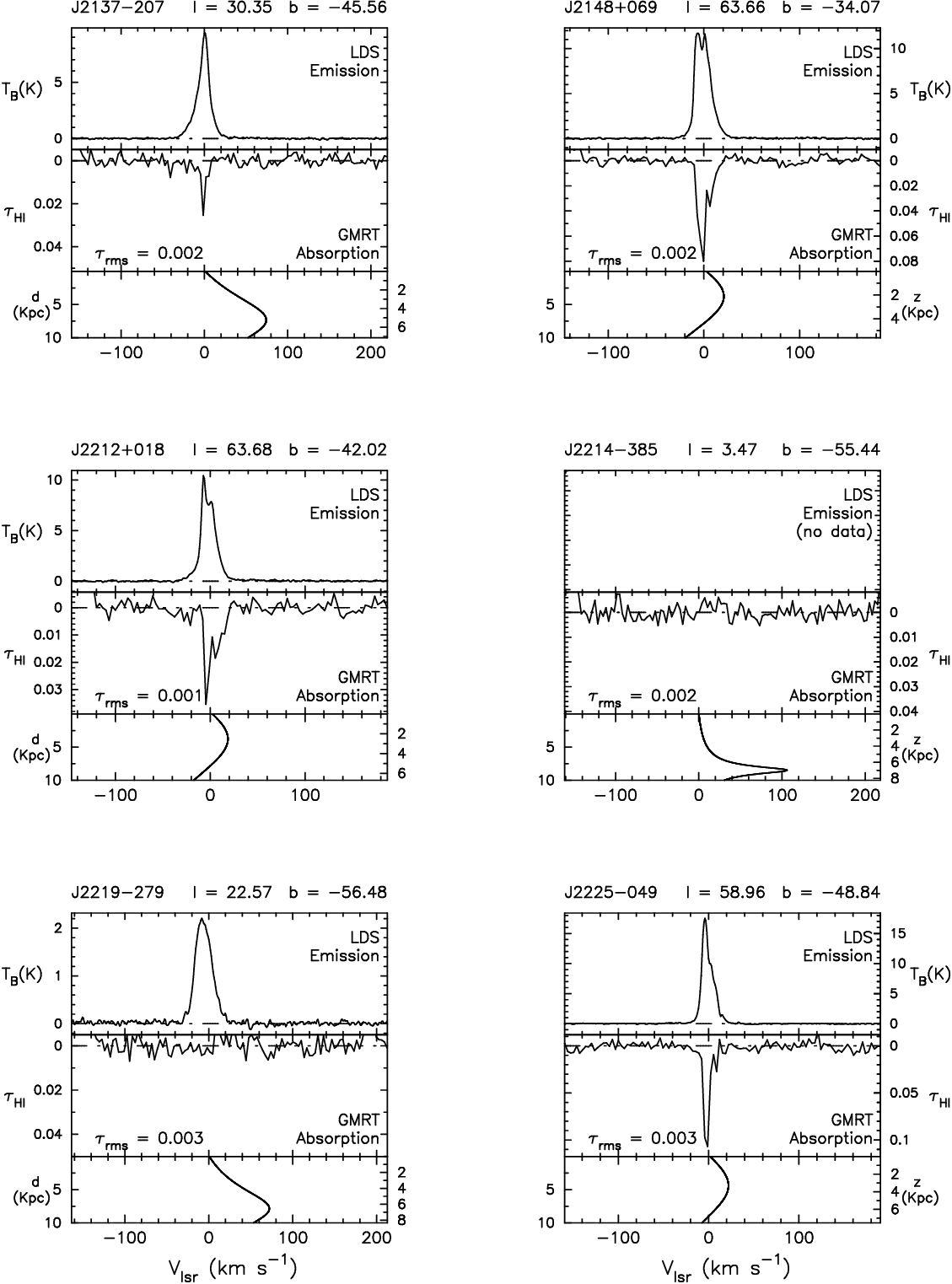}}
\end{figure}

\clearpage

\begin{figure}
\vspace{1.0cm}
{\centering\includegraphics[height=18.5cm,width=12cm]{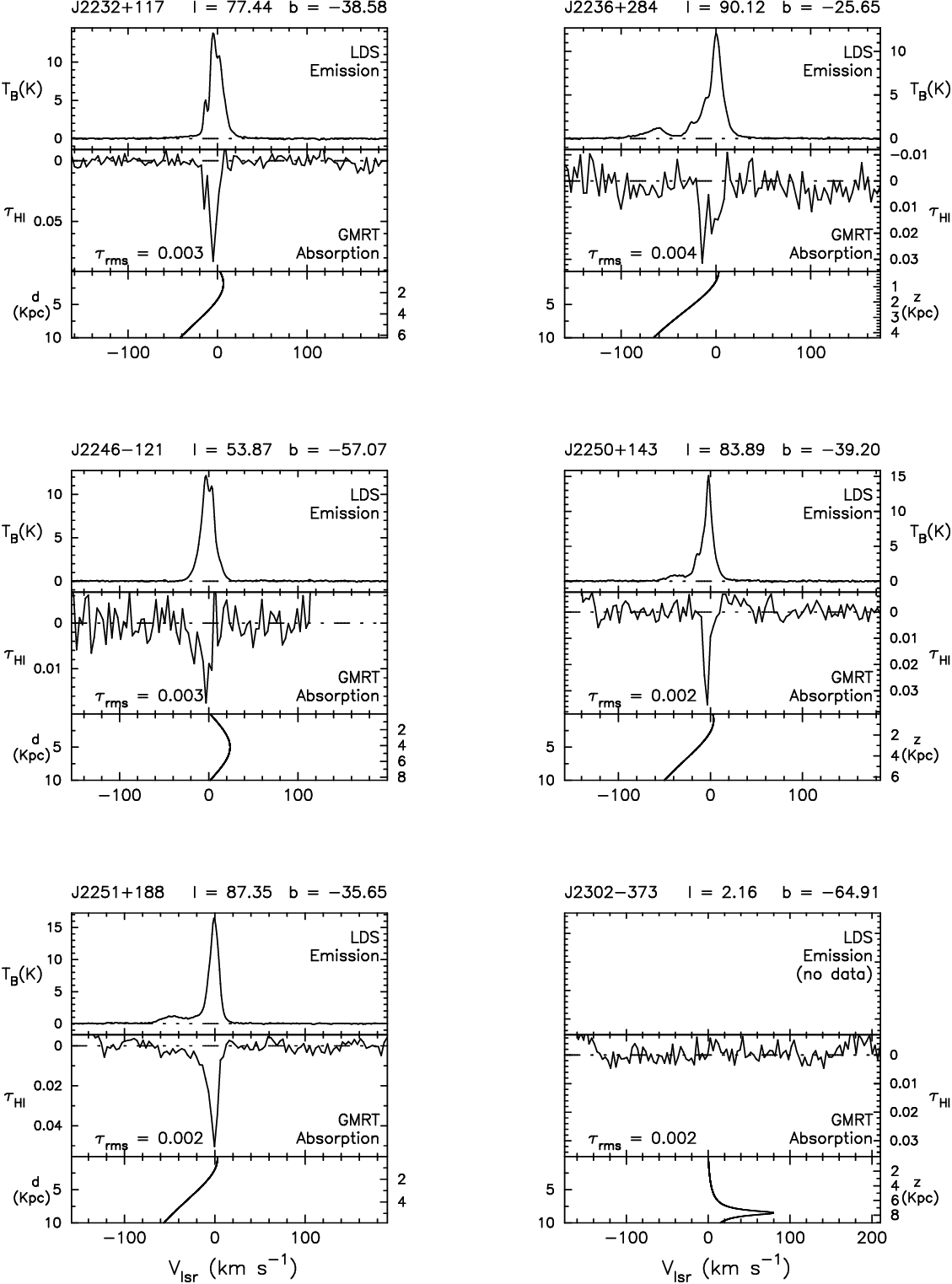}}
\end{figure}

\clearpage

\begin{figure}
\vspace{1.0cm}
{\centering\includegraphics[height=5.22cm,width=12cm]{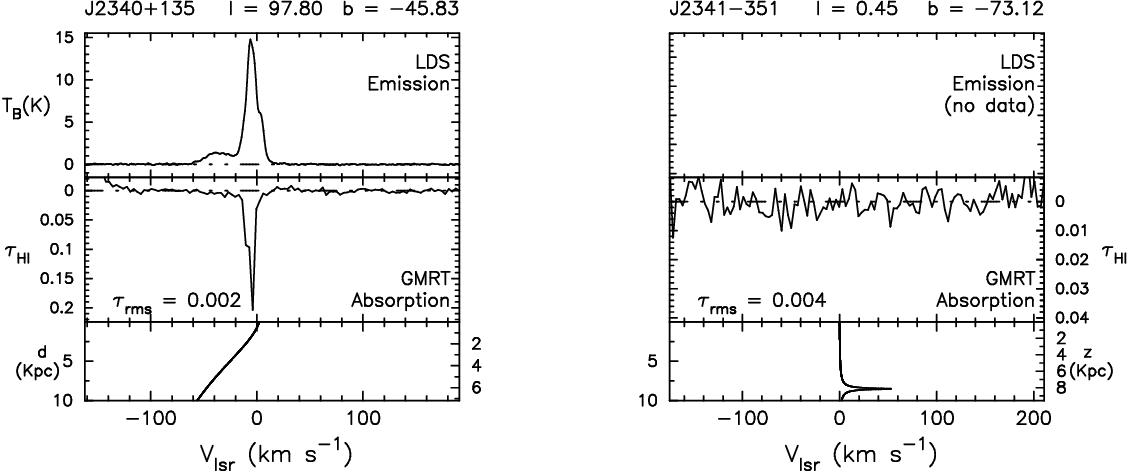}}
\end{figure}

\clearpage

\noindent \textbf {\large{Appendix C}}
\vspace{\baselineskip} \\
In this Appendix, we list the discrete HI features identified
from the high latitude Galactic HI absorption survey using the GMRT and
from the HI emission data along the respective lines of sight from the
Leiden-Dwingeloo survey of Galactic neutral hydrogen.
We have used the HI emission and absorption data to estimate the spin
temperature of the absorbing gas. \\

In the following table, the first column lists the name of
the background source, towards which the absorption was measured using the
GMRT. Columns 2, 3 and 4 list the peak optical depth, the
mean LSR velocity and the FWHM respectively of discrete components identified
using Gaussian fitting.
The value of FWHM is deconvolved for a channel width of 3.26 km s$^{-1}$. The
unresolved lines are marked with a "--" . Columns 5, 6 and 7 list the same for the
HI emission profile along the same line of sight, obtained from the
Leiden-Dwingeloo survey of Galactic neutral hydrogen (Hartman \& Burton, 1995)
The formal 1$\sigma$ errors estimated in the last digit of the fitted parameters are given
within brackets. 
Column 8 lists the spin temperature, calculated using the
absorption and emission data. Column 2 also lists the 3$\sigma$ optical depth limit obtained from the
HI absorption profiles along the respective directions.
\\

\begin{table}[ht]
\begin{small}
\begin{tabular}{l l l l l l l l l} \hline \hline
  & \multicolumn{3}{c}{{\bf HI Absorption} (GMRT)} & &\multicolumn{3}{c}{{\bf HI Emission} (LDS)}
  &   \\ \cline{2-4} \cline{6-8}
  &             &             &             &      &    &             &      &              \\
Source & $\tau_{\scriptscriptstyle{\rm{HI}}}$  &  V$_{\scriptscriptstyle{\rm{lsr}}}$  &
     FWHM     & &T$_{\scriptscriptstyle{\rm{B}}}$    & V$_{\scriptscriptstyle{\rm{lsr}}}$   &
     FWHM & T$_{\scriptscriptstyle{\rm{S}}}$   \\
  &            &(km s$^{-1}$)&(km s$^{-1}$)&& (K)      &(km s$^{-1}$)&(km s$^{-1}$)& (K)   \\ \hline
          &             &         &         &         &        &           &              &           \\
J0010--418& $<$ 0.006   &         &         &         &        &   No data &              &           \\
          &             &         &         &         &        &           &              &           \\
J0022+002 & $<$ 0.009   &         &         &         &2.5(2)  &--1.1(2)   &6.2(4)        &$>$278     \\
\hline \hline
\end{tabular}
\end{small}
\end{table}

\begin{table}
\begin{small}
\begin{tabular}{l l l l l l l l l} \hline \hline
     & \multicolumn{3}{c}{{\bf HI Absorption} (GMRT)} & &\multicolumn{3}{c}{{\bf HI Emission} (LDS)} & \\ \cline{2-4} \cline{6-8}
         &             &             &             &    &      &             &      &              \\
Source   & $\tau_{\scriptscriptstyle{\rm{HI}}}$  &  V$_{\scriptscriptstyle{\rm{lsr}}}$  &    FWHM  &   & T$_{\scriptscriptstyle{\rm{B}}}$    & V$_{\scriptscriptstyle{\rm{lsr}}}$   & FWHM & T$_{\scriptscriptstyle{\rm{S}}}$   \\
          &             &(km s$^{-1}$)&(km s$^{-1}$)& &(K)      &(km s$^{-1}$)&(km s$^{-1}$)& (K)     \\ \hline
          &             &           &        &     &          &            &          &           \\
J0029+349 & $<$ 0.006   &           &        &     & 8.27(7)  & +2.7(1)    & 10.9(2)  & $>$1378   \\
          & 0.193(5)    & --2.02(6) & 3.3(2) &     & 16.1(2)  & --3.68(1)  & 5.11(5)  &   29      \\
          &             &           &        &     & 2.58(7)  & --12.68(5 )& 4.5(1)   & $>$430    \\
\hline \hline \end{tabular}
\end{small}
\end{table}

\begin{table}
\begin{small}
\begin{tabular}{l l l l l l l l l} \hline \hline
         & \multicolumn{3}{c}{{\bf HI Absorption} (GMRT)} & &\multicolumn{3}{c}{{\bf HI Emission} (LDS)}    &   \\ \cline{2-4} \cline{6-8}
         &             &             &             &          &             &      &        &      \\
Source   &$\tau_{\scriptscriptstyle{\rm{HI}}}$  &  V$_{\scriptscriptstyle{\rm{lsr}}}$  &    FWHM  &   & T$_{\scriptscriptstyle{\rm{B}}}$    & V$_{\scriptscriptstyle{\rm{lsr}}}$   & FWHM & T$_{\scriptscriptstyle{\rm{S}}}$   \\
          &             &(km s$^{-1}$)&(km s$^{-1}$)& &(K)      &(km s$^{-1}$)&(km s$^{-1}$)& (K)   \\ \hline
          &             &           &        &     &          &            &          &           \\
J0237+288 & $<$ 0.009   &           &        &     & 7(1)     & +4.5(9)    & 8.7(8)   & $>$778    \\
          &             &           &        &     & 14(2)    & +3.25(7)   & 3.8(1)   & $>$280    \\
          & 0.347(7)    & --2.12(5) & 4.7(2) &     & 21(2)    & --0.01(9)  & 4.9(2)   & 72        \\
\hline \hline \end{tabular}
\end{small}
\end{table}

\clearpage

\begin{table}[ht]
\begin{small}
\begin{tabular}{l l l l l l l l l} \hline \hline
         & \multicolumn{3}{c}{{\bf HI Absorption} (GMRT)} & &\multicolumn{3}{c}{{\bf HI Emission} (LDS)}
         &   \\ \cline{2-4} \cline{6-8}
         &             &             &             &        &  &             &      &              \\
Source   &$\tau_{\scriptscriptstyle{\rm{HI}}}$  &  V$_{\scriptscriptstyle{\rm{lsr}}}$  &    FWHM &    & T$_{\scriptscriptstyle{\rm{B}}}$    & V$_{\scriptscriptstyle{\rm{lsr}}}$   & FWHM & T$_{\scriptscriptstyle{\rm{S}}}$   \\
          &             &(km s$^{-1}$)&(km s$^{-1}$)&& (K)      &(km s$^{-1}$)&(km s$^{-1}$)& (K)   \\ \hline
          &             &           &         &   &          &           &          &            \\
J0336+323 & 2.62(1)     & +6.698(8) & 3.7(3)  &   & 33.7(4)  & +8.36(4)  & 5.57(4)  & 36         \\
          & $<$0.012    &           &         &   & 9.8(2)   & +5.30(6)  & 15.3(2)  & $>$817     \\
          &             &           &         &   & 33.2(5)  & +4.42(3)  & 4.96(3)  & $>$36      \\
\hline \hline \end{tabular}
\end{small}
\end{table}

\clearpage

\begin{table}[ht]
\begin{small}
\begin{tabular}{l l l l l l l l l} \hline \hline
         & \multicolumn{3}{c}{{\bf HI Absorption} (GMRT)} & &\multicolumn{3}{c}{{\bf HI Emission} (LDS)}
         &   \\ \cline{2-4} \cline{6-8}
         &             &             &             &    &      &             &      &              \\
source   &$\tau_{\scriptscriptstyle{\rm{HI}}}$  &  V$_{\scriptscriptstyle{\rm{lsr}}}$  &    FWHM  &   & T$_{\scriptscriptstyle{\rm{B}}}$    & V$_{\scriptscriptstyle{\rm{lsr}}}$   & FWHM & T$_{\scriptscriptstyle{\rm{S}}}$   \\
          &             &(km s$^{-1}$)&(km s$^{-1}$)& &(K)      &(km s$^{-1}$)&(km s$^{-1}$)& (K)   \\ \hline
          &             &            &          &   &         &            &         &             \\
J0424+020 & $<$ 0.012   &            &          &   & 8.9(5)  & +11.92(5)  & 3.8(1)  & $>$741      \\
          & 0.523(5)    & +5.63(3)   & 5.4(7)   &   & 12.4(8) & +7.96(9)   & 12.5(3) & 30          \\
          &             &            &          &   & 5.8(7)  & +6.93(3)   & 3.1(2)  & $>$483      \\
\hline \hline \end{tabular}
\end{small}
\end{table}

\clearpage

\begin{table}[ht]
\begin{small}
\begin{tabular}{l l l l l l l l l} \hline \hline
         & \multicolumn{3}{c}{{\bf HI Absorption} (GMRT)} & &\multicolumn{3}{c}{{\bf HI Emission} (LDS)}
         &   \\ \cline{2-4} \cline{6-8}
         &             &             &             &     &     &             &      &              \\
Source   &$\tau_{\scriptscriptstyle{\rm{HI}}}$  &  V$_{\scriptscriptstyle{\rm{lsr}}}$  &    FWHM  &   & T$_{\scriptscriptstyle{\rm{B}}}$    & V$_{\scriptscriptstyle{\rm{lsr}}}$   & FWHM & T$_{\scriptscriptstyle{\rm{S}}}$   \\
         &             &(km s$^{-1}$)&(km s$^{-1}$)& &(K)      &(km s$^{-1}$)&(km s$^{-1}$)& (K)   \\ \hline
          &             &            &          &   &         &            &         &             \\
J0609--157& $<$0.021    &            &          &   & 6.7(2)  & +3.10(2)   & 2.97(6) & $>$319      \\
          &             &            &          &   & 18.6(3) & +7.057(9)  & 3.71(4) & $>$885      \\
          & 0.432(9)    & +10.51(8)  & 7.1(2)   &   & 20.3(4) & +8.37(5)   & 10.2(1) & 58          \\
\hline \hline \end{tabular}
\end{small}
\end{table}

\clearpage

\begin{table}[ht]
\begin{small}
\begin{tabular}{l l l l l l l l l} \hline \hline
         & \multicolumn{3}{c}{{\bf HI Absorption} (GMRT)} & &\multicolumn{3}{c}{{\bf HI Emission} (LDS)}
         &   \\ \cline{2-4} \cline{6-8}
         &             &             &             &    &      &             &      &              \\
Source   &$\tau_{\scriptscriptstyle{\rm{HI}}}$  &  V$_{\scriptscriptstyle{\rm{lsr}}}$  &    FWHM  &   & T$_{\scriptscriptstyle{\rm{B}}}$    & V$_{\scriptscriptstyle{\rm{lsr}}}$   & FWHM & T$_{\scriptscriptstyle{\rm{S}}}$   \\
         &             &(km s$^{-1}$)&(km s$^{-1}$)&& (K)      &(km s$^{-1}$)&(km s$^{-1}$)& (K)   \\ \hline
          &             &            &          &   &         &            &         &             \\
J0842+185 & $<$ 0.012   &            &          &   & 0.21(2) & +29(2)     & 21(4)   & $>$18       \\
          &             &            &          &   & 3.54(2) & +0.3(1)    & 25.5(3) & $>$295      \\
          &             &            &          &   & 0.84(5) & --14.5(1)  & 4.4(3)  & $>$70       \\
\hline \hline \end{tabular}
\end{small}
\end{table}

\clearpage

\begin{table}[ht]
\begin{small}
\begin{tabular}{l l l l l l l l l} \hline \hline
         & \multicolumn{3}{c}{{\bf HI Absorption} (GMRT)} & &\multicolumn{3}{c}{{\bf HI Emission} (LDS)}    &   \\ \cline{2-4} \cline{6-8}
         &             &             &       &      &          &             &      &              \\
Source   &$\tau_{\scriptscriptstyle{\rm{HI}}}$  &  V$_{\scriptscriptstyle{\rm{lsr}}}$  &    FWHM  &   & T$_{\scriptscriptstyle{\rm{B}}}$    & V$_{\scriptscriptstyle{\rm{lsr}}}$   & FWHM & T$_{\scriptscriptstyle{\rm{S}}}$   \\
         &             &(km s$^{-1}$)&(km s$^{-1}$)&& (K)      &(km s$^{-1}$)&(km s$^{-1}$)& (K)   \\ \hline
          &             &            &          &   &         &            &         &             \\
J1119--030& 0.034(3)    & --8.2(3)   & 3.7(4)   &   & 3.1(3)  & --3(1)     & 14(1)   &  93         \\
          & $<$ 0.006   &            &          &   & 6.4(6)  & +5.3(1)    & 9.7(5)  & $>$1067     \\
          &             &            &          &   & 8.6(2)  & +16.70(2)  & 4.93(8) & $>$1433     \\
\hline \hline \end{tabular}
\end{small}
\end{table}

\clearpage

\begin{table}[ht]
\begin{small}
\begin{tabular}{l l l l l l l l l} \hline \hline
          & \multicolumn{3}{c}{{\bf HI Absorption} (GMRT)} & &\multicolumn{3}{c}{{\bf HI Emission} (LDS)}
          &   \\ \cline{2-4} \cline{6-8}
          &             &             &             &          &             &      &       &       \\
Source    &$\tau_{\scriptscriptstyle{\rm{HI}}}$  &  V$_{\scriptscriptstyle{\rm{lsr}}}$  &    FWHM    & & T$_{\scriptscriptstyle{\rm{B}}}$    & V$_{\scriptscriptstyle{\rm{lsr}}}$   & FWHM & T$_{\scriptscriptstyle{\rm{S}}}$   \\
          &             &(km s$^{-1}$)&(km s$^{-1}$)&& (K)      &(km s$^{-1}$)&(km s$^{-1}$)& (K)   \\ \hline
          &             &            &          &   &         &            &         &             \\
J1257--319& 0.108(6)    & --16.0(2)  & --       &   &         & No data    &         &             \\
          &             &            &          &   &         &            &         &             \\
J1316--336& 0.042(9)    & --8.5(7)   & 4(1)     &   &         & No data    &         &             \\
\hline \hline \end{tabular}
\end{small}
\end{table}

\clearpage

\begin{table}[ht]
\begin{small}
\begin{tabular}{l l l l l l l l l} \hline \hline
          & \multicolumn{3}{c}{{\bf HI Absorption} (GMRT)} & & \multicolumn{3}{c}{{\bf HI Emission}
(LDS)}    &   \\ \cline{2-4} \cline{6-8}
          &             &             &             &          &             &      &      &        \\
Source    &$\tau_{\scriptscriptstyle{\rm{HI}}}$  &  V$_{\scriptscriptstyle{\rm{lsr}}}$  &    FWHM  &   &            T$_{\scriptscriptstyle{\rm{B}}}$    & V$_{\scriptscriptstyle{\rm{lsr}}}$   & FWHM &                      T$_{\scriptscriptstyle{\rm{S}}}$   \\
          &  &(km s$^{-1}$)&(km s$^{-1}$)&& (K)      &(km s$^{-1}$)&(km s$^{-1}$)& (K)   \\ \hline
          &             &            &          &   &         &            &         &             \\
J1506+375 & $<$ 0.009   &            &          &   & 0.28(2) & --53(1)    & 33(3)   & $>$33       \\
          &             &            &          &   & 0.98(3) & --07.1(5)  & 31(1)   & $>$111      \\
          &             &            &          &   & 1.64(4) & +00.2(1)   & 8.9(3)  & $>$177      \\
\hline \hline \end{tabular}
\end{small}
\end{table}

\clearpage

\begin{table}[ht]
\begin{small}
\begin{tabular}{l l l l l l l l l} \hline \hline
          & \multicolumn{3}{c}{{\bf HI Absorption} (GMRT)}& & \multicolumn{3}{c}{{\bf HI Emission}
 (LDS)}   &   \\ \cline{2-4} \cline{6-8}
          &             &             &             &          &             &      &    &          \\
Source    & $\tau_{\scriptscriptstyle{\rm{HI}}}$  &  V$_{\scriptscriptstyle{\rm{lsr}}}$  &    FWHM  &              & T$_{\scriptscriptstyle{\rm{B}}}$    & V$_{\scriptscriptstyle{\rm{lsr}}}$   & FWHM
          & T$_{\scriptscriptstyle{\rm{S}}}$   \\
          & &(km s$^{-1}$)&(km s$^{-1}$)&& (K)      &(km s$^{-1}$)&(km s$^{-1}$)& (K)   \\ \hline
          &             &            &          &   &         &            &         &             \\
J1557--000& 0.371(8)    & +0.34(7)   & 5.4(1)   &   & 28.0(3) & --1.19(4)  & 6.88(6) & 90          \\
          & $<$ 0.012   &            &          &   & 5.1(3)  & +0.2(6)    & 22.6(5) & $>$425      \\
          &             &            &          &   & 10.5(2) & +1.37(3)   & 2.6(2)  & $>$875      \\
\hline \hline \end{tabular}
\end{small}
\end{table}

\clearpage

\begin{table}[ht]
\begin{small}
\begin{tabular}{l l l l l l l l l} \hline \hline
          & \multicolumn{3}{c}{{\bf HI Absorption} (GMRT)} & &\multicolumn{3}{c}{{\bf HI Emission} (LDS)}
          &   \\ \cline{2-4} \cline{6-8}
          &             &             &             &     &     &             &      &              \\
Source    &$\tau_{\scriptscriptstyle{\rm{HI}}}$  &  V$_{\scriptscriptstyle{\rm{lsr}}}$  &    FWHM   &
          & T$_{\scriptscriptstyle{\rm{B}}}$    & V$_{\scriptscriptstyle{\rm{lsr}}}$   & FWHM
          & T$_{\scriptscriptstyle{\rm{S}}}$   \\
          &   &(km s$^{-1}$)&(km s$^{-1}$)& &(K)      &(km s$^{-1}$)&(km s$^{-1}$)& (K)   \\ \hline
          &             &            &          &   &         &            &         &             \\
J1737+063 & 0.258(7)    & +2.63(9)   & 6.0(2)   &   & 19.8(1) & +1.62(3)   & 4.65(4) & 87          \\
          & $<$ 0.015   &            &          &   & 14.6(1) & +6.02(3)   & 4.41(5) & $>$973      \\
          &             &            &          &   & 4.69(6) & +6.8(1)    & 26.1(3) & $>$313      \\
\hline \hline \end{tabular}
\end{small}
\end{table}

\clearpage

\begin{table}[ht]
\begin{small}
\begin{tabular}{l l l l l l l l l} \hline \hline
          & \multicolumn{3}{c}{{\bf HI Absorption} (GMRT)} & & \multicolumn{3}{c}{{\bf HI Emission}
 (LDS)}   &   \\ \cline{2-4} \cline{6-8}
          &             &             &             &    &      &             &      &              \\
Source    &$\tau_{\scriptscriptstyle{\rm{HI}}}$  &  V$_{\scriptscriptstyle{\rm{lsr}}}$  &    FWHM  &
          & T$_{\scriptscriptstyle{\rm{B}}}$    & V$_{\scriptscriptstyle{\rm{lsr}}}$   & FWHM
          & T$_{\scriptscriptstyle{\rm{S}}}$   \\
          &    &(km s$^{-1}$)&(km s$^{-1}$)&& (K)      &(km s$^{-1}$)&(km s$^{-1}$)& (K)   \\ \hline
          &             &            &          &   &         &            &         &             \\
J2009+724 & 0.20(2)     & --5.2(1)   & --       &   & 22.8(1) & --5.11(1)  & 9.54(5) & 126         \\
          & $<$ 0.009   &            &          &   & 7.0(1)  & --10.1(2)  & 29.2(3) & $>$778      \\
          &             &            &          &   & 3.4(1)  & --22.23(8) & 8.3(3)  & $>$378      \\
\hline \hline \end{tabular}
\end{small}
\end{table}

\clearpage

\begin{table}[ht]
\begin{small}
\begin{tabular}{l l l l l l l l l} \hline \hline
          & \multicolumn{3}{c}{{\bf HI Absorption} (GMRT)} & & \multicolumn{3}{c}{{\bf HI Emission}
 (LDS)}   &   \\ \cline{2-4} \cline{6-8}
          &             &             &             &     &      &             &      &              \\
Source    & $\tau_{\scriptscriptstyle{\rm{HI}}}$  &  V$_{\scriptscriptstyle{\rm{lsr}}}$  &    FWHM  &
          & T$_{\scriptscriptstyle{\rm{B}}}$    & V$_{\scriptscriptstyle{\rm{lsr}}}$   & FWHM
          & T$_{\scriptscriptstyle{\rm{S}}}$   \\
          &  &(km s$^{-1}$)&(km s$^{-1}$)&&  (K)      &(km s$^{-1}$)&(km s$^{-1}$)& (K)   \\ \hline
          &             &            &          &   &         &            &         &             \\
J2212+018 & $<$ 0.003   &            &          &   & 0.08(1) & +35(8)     & 109(12) & $>$27       \\
          & 0.016(1)    & +7.0(9)    & 14(2)    &   & 4.5(1)  & +0.32(8)   & 11.2(3) & 284         \\
          &             &            &          &   & 3.4(1)  & --1.7(1)   & 24.2(4) & $>$1133     \\
\hline \hline \end{tabular}
\end{small}
\end{table}

\clearpage

\begin{table}[ht]
\begin{small}
\begin{tabular}{l l l l l l l l l} \hline \hline
          & \multicolumn{3}{c}{{\bf HI Absorption} (GMRT)} & &\multicolumn{3}{c}{{\bf HI Emission}
 (LDS)}   &   \\ \cline{2-4} \cline{6-8}
          &             &             &        &     &          &             &      &              \\
Source    & $\tau_{\scriptscriptstyle{\rm{HI}}}$  &  V$_{\scriptscriptstyle{\rm{lsr}}}$  &    FWHM   &
          & T$_{\scriptscriptstyle{\rm{B}}}$    & V$_{\scriptscriptstyle{\rm{lsr}}}$   & FWHM
          & T$_{\scriptscriptstyle{\rm{S}}}$   \\
          &  &(km s$^{-1}$)&(km s$^{-1}$)&& (K)      &(km s$^{-1}$)&(km s$^{-1}$)& (K)   \\ \hline
          &             &            &          &   &         &            &         &             \\
J2251+188 & 0.041(4)    & --0.3(3)   &7.3(9)    &   & 14.74(8)& --0.27(2)  & 10.51(6)& 367         \\
          & $<$ 0.006   &            &          &   & 1.88(8) & --7.5(4)   & 30.0(9) & $>$313      \\
          & 0.015(3)    & --9(3)     & 22(3)    &   & 1.47(8) & --8.6(1)   & 5.4(3)  & 99          \\
\hline \hline \end{tabular}
\end{small}
\end{table}

\clearpage



\label{lastpage}

\end{document}